# Evaporation-driven electrokinetic energy conversion: critical review, parametric analysis and perspectives

Andriy Yaroshchuk[a,b]

[a]ICREA, pg. L.Companys 23, 08010, Barcelona, Spain

[b]Department of Chemical Engineering, Universitat Politècnica de Catalunya, av. Diagonal 647, 08028, Barcelona, Spain

## Abstract

Energy harvesting from evaporation has become a "hot" topic in the last couple of years. Researchers have speculated on several possible mechanisms. Electrokinetic energy conversion is the least hypothetical one. The basics of pressure-driven electrokinetic phenomena of streaming current and streaming potential have long been established. The regularities of evaporation from porous media are also well known. However, "coupling" of these two classes of phenomena has not, yet, been seriously explored. In this critical review, we will recapitalize and combine the available knowledge from these two fields to produce a coherent picture of electrokinetic electricity generation during evaporation from (nano)porous materials. For illustration, we will consider several configurations, namely, single nanopores, arrays of nanopores, systems with reduced area of electrokinetic-conversion elements and devices with side evaporation from thin nanoporous films. For the latter (practically the only one studied experimentally), we will formulate a simple model describing correlations of system performance with such principal parameters as the nanoporous-layer length, width and thickness as well as the pore size, pore-surface hydrophilicity, effective zeta-potential and electric conductivity in nanopores. These correlations will be qualitatively compared with experimental data available in the literature. We will see that experimental data not always are in agreement with the model predictions, which may be due to simplifying model assumptions but also because the mechanisms are different from the classical electrokinetic energy conversion. In particular, this concerns the mechanisms of conversion of evaporation-driven ion streaming currents into electron currents in external circuits. We will also formulate directions of future experimental and theoretical studies that could help clarify these issues.

## Introduction

Following the seminal work by Osterle [1], electrokinetic (EK) conversion of mechanical energy to electricity has been studied (both theoretically and experimentally) for decades [2–9]. Its major features have been established theoretically and confirmed experimentally. However, from an application-oriented point of view, it remained unclear what advantages this process could have with regard to the very well-established conventional methods such as electromagnetic. The advent of evaporation-driven electrokinetic energy harvesting changes the situation because here the mechanical-energy input arises spontaneously due to capillarity. The latter is induced by evaporation, which is a ubiquitous spontaneous process on Earth ultimately driven by solar irradiation [10]. In addition to the spontaneous evaporation (and corresponding scenarios of autonomous (low-intensity) energy supply), there are interesting opportunities of energy harvesting from waste heat. Huge amounts of waste heat are available in industry, agriculture, municipal heating, and so on [11]. Harvesting electricity from the so-called low-grade waste heat (< 100˚C) is especially difficult owing to the low efficiency of the classical Carnot cycle. On the other hand, due to the very strong dependence of saturated water-vapor pressure on temperature within this temperature range, evaporation can be intensified by an order of magnitude when temperature is raised just from 25˚C to 70˚C. Besides, industrially (and/or naturally) occurring low-humidity airstreams can be used to further enhance evaporation and energy harvesting.

This context probably explains the recent "explosion" of interest to the topic in the academic literature. After first papers published in 2017 [12,13], tens of further have followed. Many of them postulate electrokinetic conversion as the principal mechanism. Other studies suggest different mechanisms, in particular, electricity generation due to a preferential diffusion of hydrogen ions provoked by humidity gradients in certain kinds of nanoporous materials [14–16]. So far, there have been no attempts of a quantitative analysis of this mechanism, and its important details remain unclear. Therefore, it would be premature to include such studies in this critical review. On the contrary, the basics of electrokinetic energy conversion are relatively clear [2,5,6,17]. However, understanding its coupling with evaporation requires insight into several distinct fields of study such as capillarity, transport in porous media, vapor diffusion in air, aerodynamics, electrochemistry of electrode reactions and, of course, electrokinetics. This critical review is an attempt to combine the (largely available) knowledge from these fields to produce a coherent picture of capillarity-driven electrokinetic harvesting of energy from evaporation using nanoporous materials. This scope is essentially different, for example, from the extensive review of "direct electricity generation mediated by molecular interactions with low dimensional carbon materials"[18]. In that treatise, an attempt was made to overview a number of newly-proposed hypotheses concerning possible mechanisms ("phonon wind", Coulomb drag, …) of voltage/current generation due to the movement of (ion-containing) liquids along interfaces with low-dimensional carbon materials (such as graphene, for example). While such an overview has definitely been useful, the evoked mechanisms remain hypothetical and corresponding estimates are based on a number of assumptions whose scope of applicability is very difficult to evaluate for such complex systems as nanoporous electron conductors soaked with electrolyte solutions. Besides, as we will see below, energy harvesting from evaporation was observed with non-conducting as well as (semi)conducting substrates. Moreover, qualitatively, the behavior was similar in both cases.

We will start from a brief recapitulation of basics of electrokinetic energy conversion in a pressure-driven flow through a single nanopore. After that, we will consider how hydrostatic-pressure gradients arise in nanopores due to evaporation. We will demonstrate that an interplay between evaporation from the outlet and capillarity-driven viscose flow through the pore results in an optimal pore length. We will also see that semi-spherical pattern of water-vapor diffusion from outlets of single nanopores gives rise to very rapid evaporation and extremely large theoretical power densities (e.g., as much as 0.03 nW per one 20-nm pore) in electrokinetic energy conversion. It would be tempting to linearly extrapolate this to very numerous parallel nanopores. However, we will see that in arrays of multiple nanopores, the per-pore evaporation rate is drastically reduced (by several orders of magnitude) owing to external mass-transfer limitations. Nonetheless, in principle, it can remain sufficiently high for being of practical interest.

We will also address the often overlooked (in this area) issue of "streaming-current collection". In electrolyte-filled nanopores currents are transferred by ions while in external circuits they have to be transferred by electrons. Therefore, there must be electrodes in the system where ions experience (direct or indirect) discharge in electrode reactions. In addition to often unclear mechanisms of these reactions, in systems with evaporation from distinct nanopores, there is an issue of "exit" electrode location and its compatibility with the requirements of unimpeded evaporation from the pore outlets, on one hand, and of sufficient electrode area (for avoiding excessive current densities and related over-voltages), on the other. Perhaps, as a result of combination of these technical problems, no experimental results on harvesting of electricity from arrays of distinct parallel nanopores have been reported up to date.

The problem of "colocation" of one of the electrodes and of the evaporation outlet is effectively resolved in designs with distinct evaporation and EK-conversion elements (see Fig.6 below). Moreover, below we will see that due to a strongly reduced cross-sectional area of the EK-conversion element, one can potentially benefit from high capillary pressures at realistic thicknesses of such elements (see below). Despite the promise of this configuration, only two papers have been devoted to it up to date [19,20].

On the contrary, a number of groups have extensively explored experimentally "side-evaporation" configurations. In such systems, (typically) water is sucked by capillary forces into relatively thin (tens to hundreds of micrometers) sub-microporous films, is transported along them by spontaneously-arising tangential gradients of hydrostatic pressure and experiences simultaneous evaporation from their side surface (see Fig.7). Bellow, we will show that since the evaporation area is typically much larger than the film cross-section area, these pressure gradients can be quite large and can give rise to considerable voltage differences and noticeable currents. Besides, in such systems, the electrode location is less of a problem (although the mechanisms of electrode reactions in the published studies still remain unclear).

Despite the already considerable number of experimental studies, no attempts of a quantitative analysis of such systems have been published. In this critical review, we will close this gap and formulate a simple model of the process considering the film a homogeneous and isotropic porous medium characterized by a hydraulic permeability (Darcy coefficient), maximum capillary pressure, electrokinetic-charge density (or effective zeta-potential) and electric conductivity. This will enable us to explore correlations of system performance with its principal parameters such as the film geometry (length, width and thickness), average pore size, effective zeta-potential and electric conductivity. We will see that some of the predicted correlations are in agreement with the available experimental data while other are often not. We will argue that this may be related to a film cross-sectional inhomogeneity and/or anisotropy as well as broad pore-size distribution not captured by the model. Verification of correlations with some other parameters (primarily electrokinetic ones but also hydraulic permeability) has not been possible because of the lack of corresponding experimental data. For example, we will see that frequently provided information on zeta-potential is poorly defined and not really relevant, and we will outline experimental studies that could help close this gap.

This field of study is clearly a "project in progress", so the conclusions in this critical review are supported by experimental data to variable extents, some of them remaining rather conjectures. All the more so, additional studies are needed. Hopefully, this critical review will help make future studies more targeted and the design of devices for capillarity-driven energy harvesting from evaporation more rational.

## Basics of electrokinetic energy conversion

Physically, <u>streaming currents</u> arise as a result of advective movement of electrically-charged liquids close to "charged" solid/liquid interfaces in electrolyte solutions (see Fig.1).

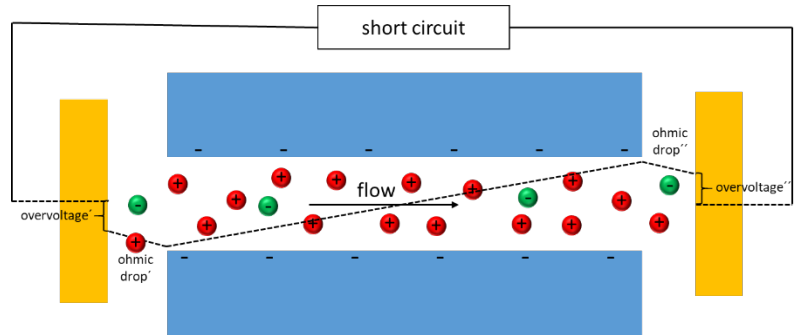

Fig.1. Schematics of streaming current.

Strictly speaking, the total electric charge of the interface region is zero, so the "charged-interface" terminology is not really rigorous (therefore, some authors refer to "electrified interfaces" [21]). Nonetheless, it is quite common and actually implies a charge separation over a certain distance (quantified by the so-called Debye or screening length). Importantly, a charge is "bound" to the surface while its "counter-charge" can move with and relative to the liquid. If there are no electric fields and ion-concentration gradients, advective movement of electrolyte solution through a nanopore with "charged" walls gives rise to a purely convective current. Its local density is equal to the product of local electric-charge density and fluid velocity. The current is the integral over the pore cross-section, and in a straight cylindrical nanopore its density is equal to

$$I_s = \frac{2}{a^2} \int_0^a v(r) \rho(r) r dr \qquad (1)$$

where $v(r)$ is the local velocity and $\rho(r)$ is the local space-charge density, $a$ is the pore radius. Expressing the space-charge distribution via electrostatic potential by Poisson equation and using Stokes equation (and the standard boundary condition of no slip on the nanopore surface) for straight cylindrical nanopores, from Eq(1) we obtain this simple expression

$$I_s = -\frac{\varepsilon\varepsilon_0}{\eta}(\zeta - \bar{\psi})\frac{dP}{dx} \qquad (2)$$

where $\varepsilon\varepsilon_0$ is the fluid dielectric constant, $\eta$ is the fluid viscosity, $\zeta$ is the electrostatic potential of pore surface (the so-called zeta-potential), $\bar{\psi}$ is the electrostatic potential averaged over the pore cross-section, $\frac{dP}{dx}$ is the gradient of hydrostatic pressure along the pore. By substituting the numbers, for water at 25°C and using a characteristic value of 25 mV for the difference between zeta-potential and average electrostatic potential, for the factor preceding the pressure gradient we obtain approximately $20 \frac{nA}{m \cdot Pa}$. Below, we will see that, for example, in single nanopores of "optimal" length experiencing water evaporation from one end, hydrostatic-pressure drops of the order of 10 MPa can occur over distances in the range of several micrometers. Taking this into account, for the streaming-current density we obtain a quite large number of $\sim 10^5 \, A/m^2$.

Eq(2) shows that streaming-current density is controlled by the difference between zeta-potential and electrostatic potential averaged over the pore cross-section (sometimes this difference is referred to as "effective zeta-potential"). Diffuse parts of electric double layers (EDL) (charged zones of liquid close to pore surfaces) in nanopores may be more or less overlapped making average electrostatic potential different from zero. The extent of the overlap increases when the screening length gets longer compared to the pore size. The screening length increases in more dilute solutions [22]. According to Eq(2), a better EDL overlap reduces streaming current[1]. Therefore, the typical use of very dilute solutions (usually, just distilled water) in the experiments does not necessarily represent an optimal scenario.

In evaporation-driven EK energy harvesting, the process is often ultimately controlled by evaporation rate. Therefore, it is useful to transform Eq(2) into an equivalent expression containing volume flux

$$I_s = \rho_{ek} J_v \tag{3}$$

where we have introduced the so-called electrokinetic-charge density (having dimensions of $A \cdot s/m^3$) according to

$$\rho_{ek} \equiv -\frac{\varepsilon \varepsilon_0}{\eta \chi}(\zeta - \bar{\psi}) \tag{4}$$

$\chi$ is the hydraulic permeability of the nanopore (proportionality coefficient between volume flux, in m/s, and negative pressure gradient, in Pa/m). Using the model of a straight cylindrical pore ($\chi = a^2/8\eta$), we obtain

$$\rho_{ek} \equiv -\frac{8\varepsilon \varepsilon_0}{a^2}(\zeta - \bar{\psi}) \tag{5}$$

By using again the characteristic value of 25 mV for the effective zeta-potential, for $a = 10\ nm$, we obtain $\rho_{ek} \sim 10^6\ A \cdot s/m^3$.

Unperturbed streaming currents occur as long as space charge is available and can be moved convectively (see Fig.1). Beyond the nanopore edges, this is not the case, so the convective current has to be either immediately "captured" by an electrode or be somehow "taken over" by a conventional (electromigration) current. In contrast to convective currents, electromigration currents are driven by voltage gradients. Accordingly, the broader are the "gaps" between the nanopore edges and the electrodes the further the observed current is from the genuine streaming current. Moreover, for observation of correct streaming currents, the electrodes must be reversible with respect to one of dominant ions present in the solution, and the exchange-current densities have to be much larger than the observed streaming-current densities. Otherwise, considerable voltage drops may occur on the electrodes, so short-circuit conditions in the external circuit do not warrant zero potential difference between the solutions just outside the electrodes (see Fig.1 for illustration). Taking as an example Ag/AgCl electrodes (used, in particular, in commercial instruments for measurements of streaming current [23], with a predominant convective movement of cations (corresponding to a negatively charged pore surface), on the Ag/AgCl cathode (the right electrode in Fig.1), a stoichiometric amount of chloride ions is released due to the transformation of AgCl to metallic silver in the electrode. The corresponding amount of electrons arrives through the external circuit. They are "generated" on the anode where chloride ions are incorporated into the AgCl layer "releasing" their electrons. Obviously, all this works well as long as the electrodes are very close to the nanopore edges, and there are no kinetic impediments for the processes of release and incorporation of chloride ions (and to the electron transfer). This occurs when the exchange-current densities are sufficiently large compared to the measured current densities. Moreover, an overwhelming majority of published studies on the EK energy harvesting from evaporation (the only exceptions being [19] where Ag/AgCl were employed and possibly [24], which used a silver paste) used non-reversible electrodes. A considerable part of studies employed carbon-based electrodes (for example, a CNT ink [25]). In this case, surface redox reactions, in principle, could occur at relatively low over-voltages as described in [26,27]. However, these reactions were studied only in acidic media, so it remains unclear if they could effectively occur in the typically pH-neutral solutions. Besides, some studies used non-carbon electrode materials (for example, copper foil [28], Au, Ag, ITO [29]) but the results were qualitatively similar to obtained with carbon-based electrodes.

---

[1] Zeta-potential can simultaneously increase in magnitude with decreasing concentration. For concentration-independent surface-charge densities, this gives rise to a monotone increase of effective zeta-potential with decreasing concentration. However, actually surface-charge densities probably decrease with concentration, so effective zeta-potentials decrease in (very) dilute solutions, too. The details depend on the pore size and the mechanisms of fixed-charge formation.

Given that all these phenomena and requirements are typically even not mentioned in the published studies on the capillarity-driven EK energy harvesting from evaporation, it is likely that the reported values of short-circuit currents (SCC) do not exactly correspond to genuine streaming currents. This should be kept in mind when comparing them with model predictions.

Observing <u>streaming potential</u> is much easier. In this mode, external circuit is open, so net electric current is zero. Accordingly, there are no potential drops in external solutions, and the electrodes can be conveniently located anywhere. True, to ensure stable measurements, they still have to be reversible. Physically, streaming potential is the voltage arising to compensate exactly the convective streaming current by an electromigration current in the opposite direction. Net current is defined as a linear superposition of convective and electromigration components

$$I = -g \frac{d\varphi}{dx} + \rho_{ek} J_v \quad (6)$$

where $g$ is the electric conductivity of solution in the nanopore. Setting this equal to zero and integrating over the pore length, we obtain

$$\Delta\varphi = \frac{\chi \rho_{ek}}{g} \Delta P \quad (7)$$

Thus, streaming-potential magnitude essentially depends on the nanopore conductance. In nanopores with (strongly) charged surfaces and (partially) overlapped diffuse parts of EDLs, electrical conductance can be strongly enhanced (due to the so-called surface conductance caused by electrostatically absorbed counterions) especially in dilute solutions [30]. Local solution conductivity is approximately proportional to ion concentrations, which are exponential functions of quasi-equilibrium electrostatic potential and can be strongly enhanced for one kind of ions (either cations or anions depending on the sign of the surface charge) The conductivity (in straight cylindrical nanopores, in (1:1) electrolytes) is given by this relationship (see the ESI).

$$g = \frac{F^2 c}{RT} \frac{2}{a^2} \int_0^a \left( D_+ exp\left(-\frac{F\psi}{RT}\right) + D_- exp\left(\frac{F\psi}{RT}\right) \right) r dr \quad (8)$$

where $D_\pm$ are ion diffusion coefficients. At larger values of quasi-equilibrium electrostatic potential, $\psi$, the conductance is roughly exponential in it, while streaming current is approximately linear (see Eq(2)). Therefore, streaming potential is a non-monotone function of surface-charge density as illustrated by Fig.2 (details of the calculations are provided in the ECI and can be found also in [31]). In this calculations, LiCl was considered as the electrolyte because this is a common salt with the least mobile single-charge cation, and it is known that electrokinetic energy conversion is more efficient for salts with less mobile counterions [9].

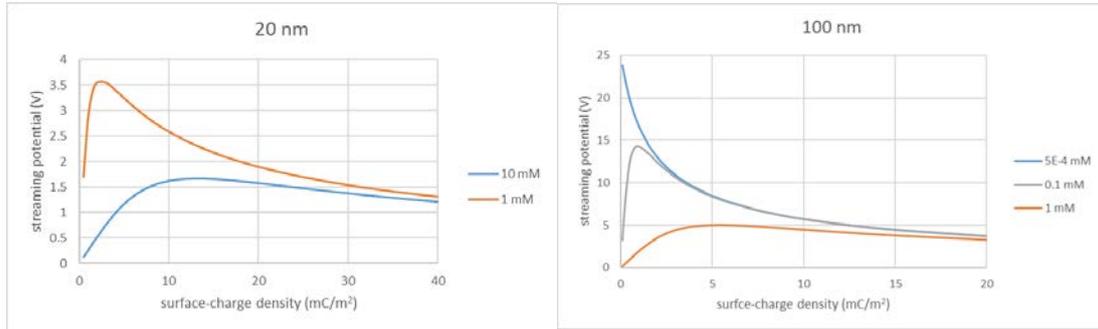

Fig.2. Streaming potential vs surface charge density: straight cylindrical pores of diameters indicated in the titles; pressure difference supposed to be equal to the maximum capillary pressure at perfect wetting (contact angle equal to zero): 14 MPa (20 nm), 2.8 MPa (100 nm); aqueous LiCl solutions of various concentrations indicated in the legends

Nanoporous materials experiencing pressure-driven electrokinetic phenomena are a particular class of Electro-Motive Forces (EMF). In (linear) EMFs, optimal power output occurs when an external-load electrical resistance is equal to their internal resistance [32]. Under these conditions, the current is half the short-circuit current (SCC) while the voltage is half the open-circuit voltage (OCV). Accordingly, the output power (equal to the product of current and voltage) under optimal external-load conditions is equal to ¼ of the product of SCC and OCV. In the systems of interest (EK convertors of mechanical energy to electricity), the SCC is the streaming current and the OCV is the streaming potential, so the optimal power output is equal to the quarter of their product. To obtain a dimensionless figure of merit for the conversion efficiency, one should scale this output on the input mechanical power. This is equal to the product of volume flow and hydrostatic-pressure difference. Taking into account that the flow rate is proportional to the pressure difference, for the dimensionless figure of merit from Eqs(3,7), we obtain

$$\Theta = \frac{\chi \rho_{ek}^2}{4g} \qquad (9)$$

where $\chi$ is the hydraulic permeability. The square of electrokinetic-charge density is proportional to the square of effective zeta-potential (see Eq(2)) while electric conductivity is roughly exponential in electrostatic potential. This can give rise to non-monotone dependencies of conversion efficiency on the surface-charge density as illustrated by Fig.3. The maxima (at realistic surface-charge densities) are more pronounced in larger nanopores. The calculations assumed negative surface charge and LiCl as the electrolyte because it has single-charge cation with the lowest mobility among common salts. From Eq(9), one can see that efficiency of EK conversion is higher in solutions of lower conductivity while Eq(8) shows that in negatively charged nanopores the latter is controlled by the mobility of cations. Replacing LiCl with KCl, for example, would reduce the efficiency by about a factor of 2. Extrapolating this trend (approximate inverse proportionality to the mobility of counterion) in HCl solutions, for example, one could expect an even much lower efficiency. However, such extrapolation may not be correct because at lower pH values many interfaces get positively charged, so $H^+$ ions become coions and the efficiency increases.

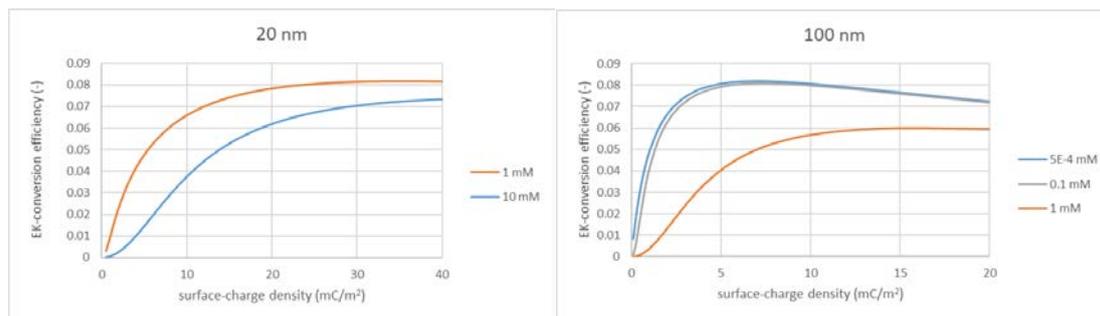

Fig.3. Electrokinetic conversion efficiency vs. surface-charge density; negatively charged surface; the diameter of cylindrical pores is indicated in the titles; LiCl solutions of concentrations indicated in the legends

One can see that for optimal combinations of parameters, the efficiency can be as high as about 8% (and stay around this level within relatively broad ranges of surface-charge densities). Remarkably, this not always occurs at excessively high surface-charge densities because with them strong surface conductance (due to electrostatically "adsorbed2 counterions) considerably reduces streaming potential, while streaming current is only marginally enhanced. This effect is more pronounced with larger nanopores and in more dilute solutions. This is a good news because surface-charge density usually decreases in dilute solutions [31,33].

Several studies reported on attempts to enhance the performance of energy harvesting from evaporation by increasing surface-charge density of porous films [34,35]. In view of our analysis, this may not necessarily be purposeful especially in larger nanopores and in very dilute solutions typically used in experiments up to date. Separate characterization of electrokinetic properties and electric conductivity is called for as a guidance on rational materials development (see below).

In systems with a genuine mechanical input (externally applied hydrostatic-pressure difference), the figure of merit of Eq(9) provides exhaustive orientation on the conversion performance. In evaporation-driven systems, of primary interest is not the efficiency but the power density *per se* while the "mechanical input" occurs spontaneously due to capillarity and can essentially depend on such properties as pore size and length, porosity, and wettability, as well as external mass-transfer conditions. The sections bellow illustrate this for four configurations: a single nanopore, infinite arrays of nanopores, systems with reduced cross-section area of EK-conversion elements and thin nanoporous films with a "side" evaporation.

## Coupled evaporation from, viscose flow and electrokinetics in a single nanopore

In this section, we will consider a single nanopore. Although this configuration is of no immediate practical interest, it will help us better understand the interplay between evaporation (and vapor diffusion) from the pore outlet and viscose flow of liquid along it. A single straight cylindrical nanopore is exposed to evaporation at one end. The other end is in electrolyte solution kept at atmospheric pressure (see Fig.4).

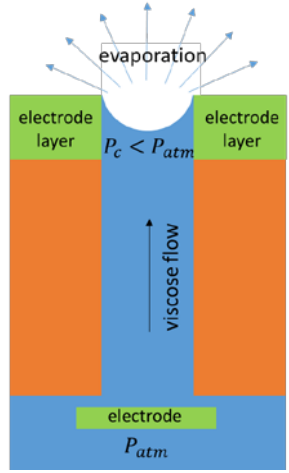

Fig.4. Schematic of coupled evaporation from and capillarity-driven viscose flow in a single nanopore.

The linear evaporation vapor flux (m/s) is given by [36]

$$J_e = \frac{4aD_v \Delta c_v V_w}{\pi a^2} \equiv \frac{4D_v \Delta P_v V_w}{\pi a RT} \tag{10}$$

where $a$ is the pore radius, $V_w$ is the molar volume of liquid water, $D_v$ is the diffusion coefficient of water vapor, $\Delta P_v$ is the difference of water-vapor pressure between the meniscus location and infinity (we assume local equilibrium between liquid and vapor at the meniscus). For simplicity we neglect the (relatively minor) effect of meniscus curvature on the saturated-vapor pressure and use the approximation of a flat circular source. Remarkably, with sufficiently small pores, the evaporation rate can be very high (in the range of cm/s) because (due to the semi-spherical diffusion pattern) it is controlled by the nanometric pore radius as a characteristic length. In systems with very numerous pores (and non-negligible porosities), the characteristic lengths for vapor diffusion are the thicknesses of stagnant layers in air, which are typically in the range of millimeters [37]. The cross-over from single nanopores to their infinite arrays will be considered in the next section.

Water lost to evaporation must be permanently replenished by a viscose flow along the pore. This is driven by the difference between atmospheric pressure and negative capillary pressure arising beneath the curved meniscus at the pore outlet. Notably, its curvature is not a "material property" but adjusts itself to the process. Indeed, the linear rate of viscose flow is given by this well-known Hagen-Poiseuille formula

$$J_{vis} = \frac{a^2 P_c}{8\eta L} \tag{11}$$

where $L$ is the nanopore length, $\eta$ is the solution viscosity, $P_c$ is the capillary pressure. Notably, for shorter nanopores, the latter is smaller than the maximum possible capillary pressure controlled by the pore radius. In steady state, the rate of viscose flow must match the evaporation rate. This can occur as long as the nanopore is short enough for the viscose flow to be driven by the capillary pressure. The latter is limited from above by the maximum capillary pressure corresponding to the largest meniscus curvature equal to the nanopore radius. Thus, there is a characteristic nanopore length corresponding to conditions where the rate of viscose flow driven by the maximum capillary pressure is equal to the evaporation rate. Using Eqs(10,11), for this length we obtain

$$L_m = \frac{\pi R T a^3 P_{cm}}{32 \eta D_v \Delta P_v V_w} \tag{12}$$

where $P_{cm}$ is the maximum capillary pressure. If the pore has the same radius along the whole length, the maximum capillary pressure is[38]

$$P_{cm} = \frac{2\sigma \cos\theta}{a} \tag{13}$$

where $\sigma$ is the water surface tension, $\theta$ is the wetting contact angle. When the nanopore becomes still longer, the maximum capillary pressure gets unable to drive the viscose flow through it, so the meniscus recedes some distance into the nanopore (keeping the same curvature), which reduces the evaporation rate and makes it equal to the reduced viscose-flow rate. The latter gets smaller because the same (maximum capillary) pressure difference is applied over a longer distance[2]. These phenomena are well known from the theory of drying of porous materials [39–42].

Thus, in shorter nanopores, the viscose-flow rate (and pressure drop) adjusts itself to the (pore-length-independent) evaporation rate while in longer nanopores the evaporation rate decreases to match the reduced rate of viscose flow. As a result, the characteristic pore length corresponds to a combination of the largest flow rate with the largest pressure drop along the nanopore. In the previous section, we have seen that streaming potential is proportional to the pressure drop while streaming current is proportional to the viscose-flow rate. Therefore, the characteristic pore length gives rise to the largest electrokinetic power (considering all other properties but the nanopore length constant) and, thus, can be considered optimal in this context.

By substituting the expression for the maximum capillary pressure (Eq(13)) to Eq(12), for the optimal length of a single nanopore we obtain

$$L_m = \frac{\pi a^2 R T \sigma \cos\theta}{16 \eta D_v \Delta P_v V_w} \tag{14}$$

For example, for a 20 nm nanopore at 25°C (assuming a 1-mm stagnant layer in air and zero ambient humidity) this length is about 2.6 μm.

As shown in the previous section, the electrical power density generated in pressure-driven flows is equal to ¼ of the product of streaming-current density and streaming potential, which for the optimal nanopore length gives

$$W_{max} \equiv \frac{1}{4} \frac{\rho_{ek}^2}{g} J_e^2 L_m \tag{15}$$

After substitution of Eq(10) and Eq(12), we obtain

$$W_{max} \equiv \frac{8\sigma \cos\theta V_w \Delta P_v D_v}{\pi R T a^2} \Theta \tag{16}$$

where the electrokinetic-conversion efficiency, $\Theta$, is given be Eq(9). The maximum power per pore is

$$\pi a^2 W_{max} \equiv \frac{8\sigma \cos\theta V_w \Delta P_v D_v}{RT} \Theta \tag{17}$$

---

[2] By using the model of in-series connection of "entrance" diffusion resistance and diffusion resistance of the part of nanopore evacuated by the receding meniscus [64], one can show that the recess length is relatively small and doesn´t noticeably reduce the length of the nanopore part remaining liquid-filled.

Remarkably, the first ("non-electrokinetic") factor in Eq(17) is independent of the nanopore size. At the same time, it strongly increases with temperature (primarily, due to the strong temperature dependence of saturated water-vapour pressure). Recall that the nanopore is assumed to always have the optimal length given by Eq(12). Therefore, the increased power density at higher temperatures is achieved with essentially shorter nanopores. In the previous section, we have seen that the dimensionless factor $\Theta$ can be made around 0.08 via adapting the surface-charge density and salt concentration to the pore size. Substituting the numbers to Eq(17), for ambient temperature (298K), assuming perfect wetting ($cos\theta = 1$), and electrokinetic efficiency factor equal to 0.08, we obtain a per-pore power of as much as ca. 0.03 $nW$, which (for example, for a nanopore of 10-nm radius) corresponds to a huge power density of about 100 kW/m²! This is still larger, for example, than the densities estimated in [43] for single nanopores under salt-concentration gradients. Such large per-pore powers are a direct consequence of the very rapid evaporation from single nanopores caused by the semi-spherical vapor-diffusion pattern. Very high linear evaporation rates (up to centimeters per second) from single nanochannels (slit like and cylindrical) were confirmed via an almost direct optical observation in [44,45]. One of these studies [44] also speculated about possible enhancement of evaporation due to water films "creeping out" of nanochannel outlets with hydrophilic surfaces and provided some indirect experimental evidence for that. Irrespective of whether such films actually occur, the experimentally observed high evaporation rates must be matched by pressure-driven solution flows along nanochannel giving rise to large streaming currents (and streaming potentials in sufficiently long nanopores).

Above, it has been implied that streaming current can be somehow "collected" by an electrode at the nanopore evaporation outlet. Given that this outlet should be exposed to air; this may be non-trivial. One of possible configuration is sketched in Fig.4. Ideally, the electrode material should be as hydrophilic as possible in order not to impair the wetting of the pore outlet. Besides, it shouldn´t be excessively thick because the streaming potential occurs only on the non-electrode part of the nanopore. On the other hand, too thin electrode layers would imply extremely high local current densities. Even assuming that there are suitable electrode reactions (most probably water splitting: from Fig.3 one can see that streaming potentials can largely exceed ca.1.2 V required for this electrode reaction to occur), potentially very high local current densities might not be possible because of sluggish electrode kinetics. Lower currents would lead to reduced power densities. The aforementioned hypothetical water films "creeping out" of hydrophilic nanopore outlets could enlarge the effective area of the "exit" electrodes and reduce the local current densities. Besides, in arrays of multiple parallel nanopores (see below), local streaming currents for each nanopore can be strongly reduced while the current (and power) densities can be kept at acceptable levels due to the parallel connection of nanopores.

## Infinite arrays of nanopores and nanoporous materials

Above, we have seen that with a 20-nm single nanopore, the per-pore electric power can be expected to be as high as ca.0.03 nW at ambient temperature and be further enhanced by about one order of magnitude if the temperature is raised to 70˚C. It is technically possible to have as many as almost $10^{14} m^{-2}$ parallel nanopores of around this size, for example, in track-etched membranes [46]. It would be tempting to extrapolate from a single nanopore to an array of $10^{14} m^{-2}$ of them just by multiplying the per-pore power by the pore density to obtain 3 kW/m². Effectively, just that was done, for example, in [43] to conclude that power densities in the process of electric energy harvesting from salinity gradients using boron nitride nanopores could be as high as several kW/m². Unfortunately, such extrapolation is correct only for pore densities that don´t give rise to power densities of practical interest while with $10^{14} m^{-2}$ of 20-nm nanopores, the extrapolation overestimates the power density by almost 4 orders of magnitude (see below), so we end up with about 0.3 W/m². This is still a decent value in this context but using such high pore densities can be additionally problematic in view of strongly increased optimal nanopore lengths. This section will demonstrate that similar (and larger) power densities can be achieved at much smaller pore densities with an additional advantage of more realistic nanopore lengths.

A recent study [36] analyzed transport phenomena described by Laplace equations (in particular, vapor diffusion in air within stagnant layers) in systems involving infinite regular arrays of "sources" (these can be, in particular, nanopore outlets in evaporation from porous materials). For infinite regular flat arrays with stagnant layers, the diffusion resistance was demonstrated to be well approximated by an in-series connection of the diffusion resistance of the stagnant layer and of an effective parallel connection of multiple nanopores. When the distance between the latter is much larger than their size, evaporation from each of the nanopores can be considered independent of the others. However, when the inter-pore spacing decreases (and becomes less than ca. tenfold of the pore size) the presence of neighboring nanopores noticeably reduces the per-pore evaporation flux. All these phenomena are captured by this simple expression for the diffusion resistance of in-series connection of a regular square array of nanopores with circular openings and a stagnant layer

$$R \approx \frac{1}{D_v}\left(\delta + h \cdot \left(\frac{h}{a} - 1.24 + \cdots\right)\right) \quad (18)$$

where $\delta$ is the thickness of the stagnant layer, $h \equiv \frac{1}{2\sqrt{N}}$ is the half-distance between the pore-outlet centers, $N$ is the pore density (number of pores per unit area). By using Eq(10) for the per-pore linear evaporation flux and Eq(18) for the diffusion resistance in an infinite array of nanopores, for the per-area linear (m/s) evaporation flux, we obtain

$$J_e = \frac{4aD_v\frac{\Delta P_v}{RT}V_w N}{1 + 4a\delta N - 2.48a\sqrt{N}} \quad (19)$$

This evaporation flux increases proportionally to the pore density as long as $4a\delta N \ll 1$. At larger pore densities, the increase becomes ever slower and reaches saturation at $4a\delta N \gg 1$. The saturation value is

$$J_e = \frac{D_v}{\delta}\frac{\Delta P_v}{RT}V_w \quad (20)$$

which is simply the 1D water-vapor diffusion flux across a stagnant layer of thickness $\delta$. For the electric output power density (the product of evaporation rate and maximum capillary pressure times EK-conversion efficiency factor) at the optimal pore length, we obtain

$$W_{max} \equiv \Theta \frac{4aD_v\frac{\Delta P_v}{RT}V_w N P_{cm}}{1 + 4a\delta N - 2.48a\sqrt{N}} \quad (21)$$

Fig.5 shows power densities calculated by using Eqs(21) for a regular square array of 20-nm cylindrical nanopores making the same assumptions as in the estimates just after Eq(17).

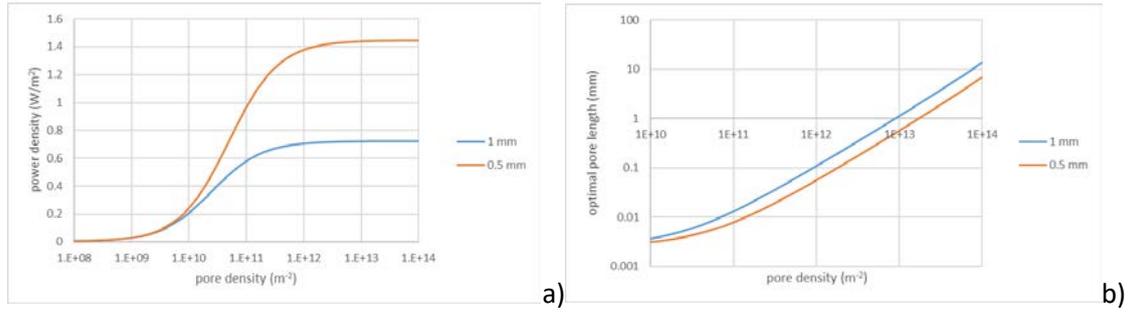

a)  b)

Fig.5. Power density (a) and optimal pore length (b) in EK energy harvesting from water evaporation: 25°C, 20 nm pore size, perfect wetting, $\Theta = 0.08$; the legends indicate the thickness of stagnant layer

Importantly, all of the pores are supposed to have an optimal length that gives rise to the pressure drop along the pore equal to the maximum capillary pressure. As long as the pore densities are low, this length is equal to the single-nanopore estimate of Eq(12). However, once the per-pore evaporation rate (and the matching viscose flow) starts to decline due to the presence of other nanopores, the optimal nanopore length has to increase to compensate for the reduced flow rate. Otherwise, streaming potential (proportional to the pressure drop) and the power density would decline. Fig.5b) shows the optimal pore length vs. pore density. This length increases considerably (up to hundreds of micrometers and even single millimeters) at higher pore densities. (Very) distant well-defined nanopores can occur, for example, in track-etched membranes[3], but for them thicknesses in the range of hundreds of micrometers are absolutely unachievable (due to technical reasons with the generation of long tracks [46]). However, the power densities are already close to their maxima at pore densities as low as ca.$10^{11}$ $m^{-2}$. Here, the optimal pore lengths can be just around 10 μm typical for track-etched membranes. Besides, this length is inversely proportional to the difference of saturated-vapor pressures, which can be strongly (by an order of magnitude) enhanced by increasing temperature from 25°C to 70°C. The optimal length decreases with temperature somewhat less (due to a partial compensation by decreasing viscosity) but still by a factor of five (see Eq(12)). Therefore, in waste-heat-harvesting applications, even somewhat larger pore densities can be compatible with realistic nanopore lengths, which can lead to a further power-density enhancement on top of the principal increase due to the strongly enhanced evaporation rate.

Alternatively, one could consider use of "conventional" nanoporous materials. However, even the $10^{14}m^{-2}$ pore density corresponds to only 4% porosity (with the pore size of 20 nm). If the porosity is still larger (which is typical for the "conventional" nanoporous materials) the optimal length (nanoporous-material thickness) further increases to tens of millimeters. This can be a handicap, in particular, in terms of per-volume power density. An expression for the optimal length (thickness) in the limit of very large pore densities is easy to obtain from the "saturation" value of evaporation flux (Eq(20)) and an expression for the viscose-flow flux taking into account finite porosity of the porous material.

$$L_m = \frac{RTa^2\gamma P_{cm}}{8\eta D_v \Delta P_v V_w}\delta \qquad (22)$$

---

[3] Their commercial nanoporous grades have much higher pore densities (around $5 \cdot 10^{13} m^{-2}$) [46] but reducing the track density down to a required value via diminishing the irradiation dose should not be a problem.

where $\gamma$ is the "active" porosity accounting for pore tortuosity. Assuming for it a relatively low (but still realistic) value of 0.1 and for the other parameters the same values as in the estimates of the optimal length made above for a single nanopore, we obtain $L_m \approx 30\delta$, which gives about 3 cm for a typical stagnant-layer thickness in air of 1 mm (at the wind velocity of 1 m/s and the characteristic dimension of 6 cm) [37]. At the same time, as discussed above, at higher temperatures (viz. waste-heat energy harvesting) the optimal thickness can become essentially smaller. Besides, the streaming-current "collection" in this configuration could be less of a problem due to strongly reduced local current densities.

Fig.5 also illustrates the importance of stagnant-layer thickness (and external mass transfer). When this thickness decreases the power density increases reciprocally. As just mentioned, typical stagnant-layer thicknesses in air under mild conditions of 1 m/s airflows are about 1 mm. Thinner layers can be expected with stronger airflows and for structures mimicking, for example, tree leaves (due to their relatively small dimensions and ability to move).

In summary, the configuration of arrays of scarce parallel nanopores can be of practical interest especially for waste-heat energy harvesting. However, current collection on the evaporation side of such systems can be challenging and remains unexplored. A configuration of relatively thick (millimeters to centimeters) layers of "conventional" nanoporous materials may also show interesting performance but this would require the use of "monoliths" with the same (nanoscale) average pores size and a relatively narrow pore-size distribution (and no cracks or other defects) across the (considerable) whole thickness. To our knowledge, this scenario also remains unexplored.

## Reduced viscose-flow cross section

Above, we have seen that in arrays of multiple nanopores, per-pore evaporation rate is strongly reduced because of external mass-transfer limitations. As a result, in 1D flows the pressure drop (and streaming-potential difference) along the pore can approach the maximum corresponding to the maximum capillary pressure only for quite long nanopores, which can be technically disadvantageous. This section considers an alternative scenario of using a smaller cross-section area for the viscose flow than for the evaporation. This approach was suggested (and partially explored experimentally) in [19]. The concept is schematically shown in Fig.6.

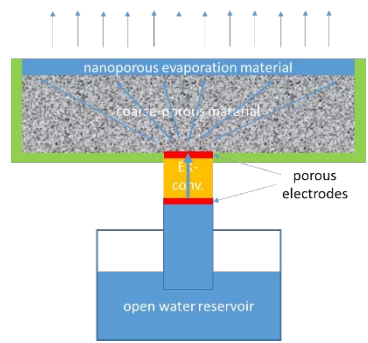

Fig.6. Schematic of EK energy harvesting with reduced EK-conversion cross-section area (not to scale)

Due to the reduced area of the "EK-conversion" element the (relatively low) linear evaporation rate can be considerably amplified in this element. Namely, the pressure drop (and streaming-potential difference) here can be strongly enhanced, while streaming current can also be relatively large due to the big evaporation area.

On the contrary, in a (not too thick) nanoporous evaporation layer, for the usually quite low "macroscopic" evaporation rates to be matched by viscose flow, only small pressure differences are needed. Thus for instance, using the classical Hagen-Poiseuille expression for the hydraulic permeance of porous medium ($a_e^2 \gamma_e / 8\eta\, L_e$, $a_e$ is the pore radius, $\gamma_e$ is the effective porosity, $L_e$ is the evaporation-layer thickness), assuming a fairly large linear evaporation rate of 1 μm/s, the layer thickness of 10 μm, the pore radius of 10 nm and the effective porosity of 30%, the pressure difference needed to match the evaporation rate is just about 20 kPa. This is very little compared to the maximum capillary pressure of about 10 MPa occurring for such a material at perfect wetting. Therefore, by adjusting the area, pore size, porosity, and thickness of the EK-conversion element, the pressure drop on it can be made practically equal to the maximum capillary pressure, which can occur under fully developed menisci in the evaporation element. However, very thin nanoporous evaporation layers of macroscopic lateral dimensions can hardly be used in practice without mechanical supports, for example, relatively coarse-porous materials. The properties of such materials (pore size, porosity, thickness) should be defined on case-by-case basis depending on other parameters of the device but negligible pressure drops on them seem feasible. Notably, in [19], a fragile nanoporous alumina membrane used for evaporation was not supported mechanically. This was possible because the EK-conversion nanofluidic element was very thin (60 μm) and had a not too small area (ca.2 mm$^2$), so the pressure drop on it was much smaller than the maximum capillary pressure and did not exceed ca.5 kPa [19]. The pressure drop on the evaporation membrane was still about 250 times lower (inversely proportionally to the ratio of the element cross-section areas), which made possible the use of (an unsupported) very fragile alumina membrane.

Neglecting the small pressure drops on the evaporation layer and the eventual supporting coarse-porous material, the whole capillary-pressure drop should occur on the nanofluidic EK-conversion element. Similarly to the case of evaporation from single nanopores, there is an optimal thickness of this element giving rise to the maximum capillary-pressure drop on it

$$L_f = \frac{a_f^2 \gamma_f}{8\eta} \left(\frac{S_f}{S_e}\right) \frac{P_{cm}}{q_e} \qquad (23)$$

where $a_f$ is the pore radius in the nanofluidic element, $\gamma_f$ is its effective porosity (accounting for the pore tortuosity), $S_f, S_e$ are the cross-section areas of the nanofluidic and evaporation elements, $q_e$ is the linear evaporation rate (expressed as corresponding volumetric flow per unit area, in m/s) occurring with menisci located just at the pore outlets. Above (see Eq(22)), we have seen that for coupled 1D evaporation and viscose flow, the optimal nanoporous-material thickness can be as large as several centimeters and more (especially, for larger pores and porosities). Eq(23) demonstrates that this optimal thickness can be considerably reduced by using EK-conversion elements of smaller cross-sections than that of the evaporation ones. Optimal combinations of thickness and extent of cross-section reduction depend on the kind of material used for the EK-conversion element. An additional advantage is the possibility of fine-tuning other properties of each material independently. Thus for instance, while smaller pores in the evaporation material are desirable because of increased capillary pressure, excessively small pores in the EK-conversion element may bring about sub-optimal performance, for example, owing to the reduction of effective zeta-potential and/or excessive surface conductance (see Eqs(2,8)). If the EK-conversion element has the optimal thickness given by Eq(23), the streaming-potential difference is given by Eq(7) with the pressure difference equal to the maximum capillary pressure. The flow rate is $q_e S_e$, and the streaming current is given by Eq(2). Taking into account the definition of electric power as a quarter of product of streaming potential (OCV) and streaming current (SCC) and scaling it on the evaporation area, for the power density, we obtain

$$W_{max} = \Theta q_e P_{cm} \qquad (24)$$

The thickness and cross-section area of nanofluidic element don´t feature explicitly but it is implied that the relationship between them satisfies Eq(23). Notably, Eq(24) is similar to Eq(33) below but predicts a maximum power density that is about 70% higher than in the "side-evaporation" configuration mostly studied experimentally up to date. However, technical implementation of systems with reduced cross-section of EK-conversion element (especially in attempts to reach high power densities) is certainly more difficult in view of sealing issues with large negative pressures, compatibility of good sealing with electrical access to the electrodes, and so on.

A related approach was put forward in [20] where pieces of cotton fabrics were used as evaporation elements. However, this configuration seems to be less efficient because considerable parts of capillary pressure can be "consumed" (not contributing to streaming potential) for the fluid delivery along relatively thin and large (for example, 7 x 7 cm$^2$) pieces of evaporation material. Besides, the pore size in the fabrics cannot be very small, so the maximum capillary pressures are relatively low.

## Electrokinetic energy harvesting from "side" evaporation

In this section, we will formulate (and qualitatively compare with published experiments) a simple model for the description of the configuration, which has been mostly studied experimentally up to date. Although some elementary expressions for basic electrokinetic phenomena of streaming potential and streaming current in this context were considered previously [16,18,28,29,47,48], no attempts to formulate a consistent model have been made. The configuration is schematically represented in Fig.7.

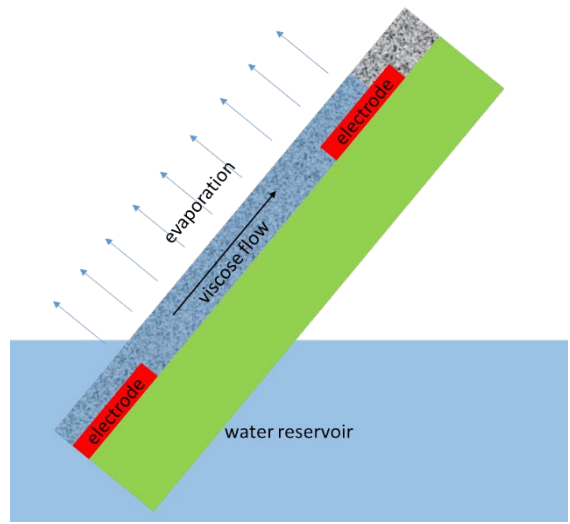

Fig.7. Schematic of systems with "side" evaporation (not to scale).

A thin film of a nanoporous material (either supported by a solid substrate or free-standing) is immersed with one extremity in an electrolyte solution. The liquid is sucked into the pores by capillary forces. Simultaneously, the solvent (typically, water) evaporates predominantly from the film side surface having the largest area. There are, at least, two electrode stripes (typically beneath the film): one located close to the immersed film extremity and another situated at a certain distance along the film length (the direction of capillary imbibition). The electrode materials have often been carbon- or silver-paste-based (see below) but the mechanisms of (and, generally, even the need for) electrode reactions have not been recognized. In some studies, there have been additional intermediate electrodes used to monitor the OCV distribution along the film.

The simple model below assumes that the porous medium is isotropic and its specific properties (porosity, pore size, electrokinetic-charge density, electric conductivity) are the same across the film thickness. This is important to note in view of some experimental studies using clearly anisotropic materials (see, for example, [24]). In experiments, the film thicknesses have been much smaller than the other dimensions. The evaporation from the exposed film surface has to be matched by a pressure-driven viscose flow. We describe this flow by Darcy law (flow rate proportional to negative hydrostatic-pressure gradient) and consider rectangular films with a length and a width that are much larger than the thickness. Due to volume conservation inside the film, a normal liquid flow has to be (approximately) matched by a flow along the film. The corresponding normal and tangential pressure gradients are commeasurable but the film thickness is much smaller than the length, so the pressure drop across the film (perpendicularly to the evaporation surface) is relatively small. In this simple model, we neglect it and consider only pressure distribution along the film. All the fluxes are assumed to be scaled on the film width. The local pressure-driven liquid flux (per unit width) along the film is

$$Q = -\chi h \frac{dP}{dx} \tag{25}$$

where $h$ is the film thickness, $\chi$ is its hydraulic permeability. Evaporation rate is assumed to be constant along the film (we disregard the dependence of saturated-vapor pressure on the menisci curvature, which, actually, changes along the film). Since the liquid permanently evaporates, the tangential volume flux is not constant but satisfies this material-balance relationship

$$\frac{dQ}{dx} = -q_e \tag{26}$$

where $q_e$ is the linear evaporation rate (evaporated volume per unit area, in $m/s$). From Eqs(25,26), we obtain

$$\frac{d^2P}{dx^2} = \frac{q_e}{\chi h} \tag{27}$$

At the "entrance" (the immersed end), the system is at atmospheric pressure (the menisci curvature is zero, this neglects the existence of external adhering solution films, see below)

$$P(0) = 0 \tag{28}$$

From the material balance, the amount of liquid "entering" (via the pressure-driven tangential flow) the porous film at the immersed end must be equal to the amount of liquid evaporating from the whole surface of the film of length, $L$ (with this, we neglect evaporation from the film edges, which is legitimate for very thin films)

$$-\chi h \frac{dP}{dx}\bigg|_{x=0} = q_e L \tag{29}$$

Integrating Eq(27) with the boundary conditions of Eqs(28,29), we obtain

$$\frac{dP}{dx} = \frac{q_e}{\chi h}(x - L) \tag{30}$$

$$P(x) = \frac{q_e}{\chi h} x \left(\frac{x}{2} - L\right) \tag{31}$$

The longer the film the larger the evaporation area and the total evaporated amount. The latter has to be matched by an ever larger pressure-driven flow along the film. Intuitively, for excessively long films this should not be possible because too high pressure differences would be required, so there must exist a maximum wet length. In other words, porous films should sooner or later dry out when moving away from the immersed end. The maximum wet length can be found from the following considerations. The tangential hydraulic flow is driven by the gradient of (negative) capillary pressure arising beneath the curved menisci at the external film surface. While moving along the film away from the immersed end, ever larger negative pressures are required to drive the viscose flow through the ever longer film segment. This negative-pressure build-up occurs due to a gradually increasing menisci curvature, which keeps growing until it reaches the maximum capillary pressure corresponding to the pore size and the wetting contact angle. Once this state is reached, the menisci curvature cannot increase anymore and menisci start to recede into the pores. The details of processes within this zone may be complex and are not quite clear but one can assume its length to be commeasurable with the film thickness. This is much smaller than the length of the fully wet zone. Therefore, in a first approximation, we will neglect evaporation from this (relatively short) partially-wet zone and assume that all the liquid evaporates from the fully wet zone having an (a priori unknown) length $L_w$. Finally, as argued above, at the end of this zone, hydrostatic pressure is equal to the maximum negative capillary pressure

$$P(L_w) = -P_{cm} \tag{32}$$

By substituting this to Eq(31), we obtain

$$L_w^2 = \frac{2\chi h P_{cm}}{q_e} \equiv \frac{a^2 h \gamma P_{cm}}{4\eta q_e} \tag{33}$$

In the right-hand side of Eq(33), we used the model of straight cylindrical capillaries for the hydraulic permeability ($\chi = a^2 \gamma/8\eta$, $\gamma$ is the film active porosity (accounting for pore tortuosity)). For the pore radius of 300 nm[4], assuming active porosity $\gamma = 0.3$, film thickness 100 µm, perfect wetting ($cos\theta = 1$) and water-evaporation rate of ca.0.7 $\mu m/s$ (this postulates the stagnant-layer thickness of 1 mm, zero relative humidity at its external boundary and saturated water-vapor pressure corresponding to 25˚C), for the fully-wet zone length, we obtain about 2 cm, which has been a typical film length in many published experiments. Actually, evaporation rates could probably be even lower than assumed above (for example, in [49] the linear evaporation rate was measured to be just about 0.07 $\mu m/s$ probably due to non-zero humidities outside the stagnant layer and lack of forced airflow) and the fully-wet lengths, accordingly, still larger (in some experiments [29] the distances between the electrodes was as large as ca.4 cm). On the other hand, in several studies the films were thinner than the assumed 100 µm, which had to give rise to shorter fully-wet zones. The existence of an optimal film length (based on qualitative considerations and experimental data) was suggested in [28,34].

In terms of electrokinetic phenomena, this system has particularities because the volume-flow rate changes along the film due to evaporation. Therefore, local streaming-current density (proportional to the flow rate according to Eq(3)) changes along the tangential coordinate. Given that the net electric-current density must be constant under these 1D conditions, there are electric-potential gradients along the film even at zero potential difference imposed between the film ends via external short-circuit (see above about electrochemical aspects of this assumption). Accordingly, the local current has both advection and migration components, so the current per unit width is

$$i = -gh\frac{d\varphi}{dx} + \rho_{ek}Q(x) \equiv -\left(hg\frac{d\varphi}{dx} + \rho_{ek}q_e(x-L)\right) \tag{34}$$

where we have substituted Eq(30) for the pressure gradient[5]. By integrating from 0 to $L$, under short-circuit conditions ($\varphi(L) = \varphi(0)$), for the "global streaming-current" per unit film width we obtain

$$i_s = \frac{1}{2}\rho_{ek}q_e L \tag{35}$$

---

[4] Information on this important parameter is typically not provided but on the published electron micrographs pores visually often look around this size or even larger.

This "streaming current" increases linearly with the film length because the total evaporated amount is proportional to it, so the tangential flow rate (and the pressure gradient) at the film "entrance" has to increase linearly with the length. The factor ½ arises because the pressure gradient linearly decreases in magnitude along the film (see Eq(30)). The per-width current is independent of the film thickness because it is proportional to the total electric charge transferred convectively, and the latter is proportional to the entrance volumetric flow. When the thickness increases, the decreasing entrance pressure gradient is exactly compensated by the increasing cross-section height, so the volumetric flow remains the same.

For the electric-potential profile along the film in the streaming-current mode (assuming the short-cut conditions between the "entrance" and the end of the film, $\varphi(L) = \varphi(0)$), we obtain

$$\varphi(x) = \varphi(0) + \frac{1}{2}\frac{\rho_{ek}}{g}\frac{q_e}{h}x(L-x) \qquad \text{streaming-current mode} \tag{36}$$

With this profile, the electric field gives rise to a migration electric current that partially compensates the "excessive" advection current at the "entrance" and enhances the "insufficient" advection current at the "exit".

If the short-cut conditions are imposed at an intermediate point, $x < L$, the corresponding "streaming current" is

$$i_s(x) = \rho_{ek}q_e\left(L - \frac{x}{2}\right) \tag{37}$$

In the <u>streaming-potential mode</u>, the net electric current is zero everywhere because the external circuit is open. From this (and Eq(34)), we obtain this expression for the local electric-potential derivative

$$\frac{d\varphi}{dx} = \frac{\rho_{ek}}{g}\frac{q_e}{h}(L-x) \tag{38}$$

The potential distribution along the film (measurable with a set of "intermediate" electrodes, see below) is

$$\varphi(x) - \varphi(0) = \frac{\rho_{ek}}{g}\frac{q_e}{h}x\left(L - \frac{x}{2}\right) \qquad \text{streaming-potential mode} \tag{39}$$

---

[5] Here we use a different notation for current than in Eqs(1-3,6,8) to stress that now the current is scaled on the film width and not on the cross-section area as previously.

One can see that the voltage gradient decreases linearly while potential increases sub-linearly. Below, we will compare these predictions with experimental data. The potential difference over the whole film is

$$\varphi(L) - \varphi(0) = \frac{1}{2}\frac{\rho_{ek}}{g}\frac{q_e}{h}L^2 \tag{40}$$

The strong increase with the length is due to the increasing total evaporated amount (hence, larger entrance gradients of both pressure and voltage) and larger potential drops over larger lengths at given entrance gradients, which gives the quadratic dependence. The inverse proportionality to the thickness occurs because in thicker films less pressure gradient at the entrance is needed to move the evaporated liquid (whose amount is independent of the thickness because evaporation occurs mostly from the external film surface).

As discussed in the section on single nanopores, the maximum harvested electric power is ¼ of the product of streaming current density and streaming potential. Eqs(37,39) show that while streaming-potential difference increases sub-linearly along the film, "streaming current" somewhat decreases with the tangential coordinate. As a result, their product has a maximum at $x = 2L/3$. Thus, the power is the largest when the upper electrode is located not at the end of the film but at an intermediate position. Nevertheless, when estimating the power density, we should scale the power on the area of the whole film (including its part "above" the electrode because without it the system would not work properly). By using Eqs(37,39), for the maximum power density (power per unit film area) occurring for the <u>optimal upper-electrode location at $x = 2L/3$</u>, we obtain

$$W_{max} = \frac{2}{27}\frac{(\rho_{ek}q_e)^2}{gh}L^2 \tag{41}$$

The power density strongly increases with the film length. Therefore, for maximizing it the films should be as long as possible, that is they should have the maximum fully-wet length given by Eq(33). By substituting it to Eq(41) we obtain

$$W_{max} = \frac{16}{27}\Theta q_e P_{cm} \tag{42}$$

The product $q_e P_{cm}$ has the dimensions of power density and can be considered a "mechanical-energy" input while the dimensionless coefficient, $16\Theta/27 \approx 0.6\Theta$, quantifies the efficiency of EK-conversion in this system. It is somewhat smaller than for single nanopores and systems with reduced EK-conversion cross-section area (see above) because the convective-current density in side-evaporation systems linearly decreases with the tangential coordinate while streaming potential increases sub-linearly. Above we have seen that factor $\Theta$ is realistically limited from above by around 0.08 even assuming optimal pore size, surface charge density and electrolyte kind. Taking the same numbers as in the above estimates of the fully-wet length (and postulating 0.08 efficiency of EK conversion), we obtain the power density of about 10 mW/m². This is in line with the largest experimental values reported up to date (see, for example, [24]) although in most cases the power densities were essentially lower. Thus for instance, they were around 150 µW/m² [50], 100 µW/m² [51], 130 µW/m² [13], 800 µW/m² [28], or 80 µW/m² [29]. Such sub-optimal performance could be primarily due to over-voltages and sluggish kinetics of electrode reactions as well as low EK-conversion efficiencies caused by insufficient surface charge-densities in very dilute electrolyte solutions used in the experiments. However, with 600-nm pores the ion concentrations could only be very low because otherwise the extent of EDL overlap in pores would not be sufficient. Further research on pore-surface modification, for example, along the lines of ref.[35] is needed to optimize the surface charge in very dilute solutions (and/or optimize the solution concentrations). Quantitative methods of in-situ electrokinetic characterization of film materials (see below) could be helpful.

Eq(42) also shows that the power density can be considerably increased by using smaller nanopores (due to increased maximum capillary pressure). Thus, for instance, with 20-nm pores (and the same values of other parameters as in the estimates above), the power density can increase up to about 0.3 W/m$^2$. True, the fully-wet porous films in this case will be essentially shorter (only about 4 mm according to Eq(33)). This may be a technical problem. Several studies have reported on the existence of solution films adhering to the external porous-film surfaces close to the solution level in the immersion reservoir. It was observed that within such zones, there are reduced electric-potential gradients along the film [25] because in such cases the liquid flows mostly outside the porous film practically without pressure gradients inside it. The length of such external adhering liquid films has been reported to be in the range of several millimeters [25] which is comparable with the maximum fully-wet zone length estimated above for 20-nm pores. Device optimization in this case could require use of narrower electrode stripes and location of the lower of them somewhat above the solution level (to have the adhering external solution film located beneath this electrode). Thicker porous films with larger porosities could also help make the fully-wet zones longer (see Eq(33)) without changing the power density (note that power density does not depend on the layer thickness, of course, as long as it is much smaller than the length).

Above we have already discussed the issue of electrode reactions. Their mechanisms in this context remain unexplored. Nonetheless, at voltage differences in excess of 1.23 V water splitting is possible [52]. Now, we will find out what voltages can be expected in the systems of interest. Substituting the expression for the maximum fully-wet film length (Eq(33)) to the relationship for the voltage difference along the whole film length[6], we obtain

$$\varphi(L_w) - \varphi(0) = \chi \frac{\rho_{ek}}{g} P_{cm} \equiv 4\Theta \frac{P_{cm}}{\rho_{ek}} \tag{43}$$

As we have seen above, factor $4\Theta$ in the right-hand side of Eq(43) is dimensionless and can be as large as about 0.3. The second factor is the ratio of maximum capillary pressure and electrokinetic charge density. With the 600-nm pores considered above as one of examples, the maximum capillary pressure can be up to ca.0.5 MPa. As discussed above, with such (relatively large) pores the dimensionless "electrokinetic" factor can be close to its theoretical maxima only in very dilute solutions (around 0.01 mM, to have an optimal EDL overlap in the pores). With overlapped EDLs, electrokinetic charge density is controlled by the surface-charge density, which for such dilute solutions is not known even by the order of magnitude. In a recent study [31], quantitative information on the surface-charge densities in 24-nm pores of a PET track-etched membrane was obtained from interpretation of simultaneous measurements of osmotic pressure and salt diffusion in milli-molar KCl and LiCl solutions. Of course (due to the extreme pore anisotropy) such membranes cannot be used in the systems with side evaporation but the results give us, at least, some orientation in terms of realistic values of surface-charge density. The latter was estimated to be around -6 mC/m$^2$ in 1.5 mM and about -10 mC/m$^2$ in 3 mM solutions of both electrolytes. The ESI shows that with these surface-charge densities (and assuming a perfect wetting of 24-nm pores) the maximum voltage drop can be estimated to be, for example, 3.4 V in 1.5 mM LiCl and 1.9 V in 1.5 mM KCl. The corresponding EK-efficiency factors ($16\Theta/27$) are 0.031 (LiCl) and 0.018 (KCl) and the expected power densities are around $0.1 \div 0.2 \ W/m^2$. Both voltages are noticeably above the water-splitting threshold (see also Fig.2), so this electrode reaction, in principle, could support the streaming-current "collection" by the electrodes. Of course, one has to keep in mind the corresponding pH changes and their possible impact on the electrokinetic phenomena. This should be explored in future studies. The possible role of water splitting is indirectly confirmed by the fact that in a system with relatively high evaporation-induced voltage (ca.2.5 V), the performance (both OCV and SCC) were practically independent of the kind of electrodes used (carbon, Au, Ag, ITO) [29].

---

[6] For the optimal location of the upper electrode at $x = 2L_w/3$, the potential difference is about 10% smaller.

These estimates also additionally illustrate the importance of electrolyte selection: both voltage and power density in LiCl are about two times larger than in KCl (see the ESI). This is because the surface conductance (reducing streaming potential, see Eq(3)) is controlled by the mobility of counterions, and the latter for Li$^+$ is about two times lower than for K$^+$. Therefore, "pure" water as the solution (used in many published studies) is not the best choice because dominant counterions probably are very mobile H$^+$ ions, in this case.

On the other hand, gradual accumulation of salt due to evaporation can be a problem in practical applications. A solution could be introducing a certain amount of salt (corresponding to an optimal concentration) to the nanoporous film during an initial system "conditioning" and using practically pure water in the further operation. Given that the linear infiltration rates (especially, at the immersed end) are typically quite high (due to the large side-evaporation area), the Péclet number (estimated with the film length) can be very large (in the order of tens to hundreds). Accordingly, the salt losses due to back diffusion from the film to the reservoir will be minor. Of course, even small amounts of ions present in the "feed" water will cause slow changes in the solution composition in the nanopores, so the system will require a periodic "conditioning" via equilibration with an "optimal" solution. However, the frequency of such "conditioning" may be sufficiently low to be compatible with certain application scenarios.

## Comparison with published experimental data

This section considers correlations between experimental data and model predictions. The experimental data are available almost exclusively for the side-evaporation configuration, alone.

### Evaporation rate

The "driving force" of this process is (water) evaporation. This has been confirmed experimentally on numerous occasions via observation of positive correlations of OCV and SCC with parameters controlling the rate of evaporation, namely, temperature, ambient relative humidity and airflow intensity. Thus for instance, the OCV considerably increased in the temperature range from 22˚C to 82˚C, which can be of interest in the context of waste heat harvesting [50]. Moreover, the SCC increased about 5 times when temperature increased from 24˚C to 55˚C in quantitative agreement with the increase in saturated-vapor pressure [24]. Further positive correlations with temperature [19,29,47] and airflow rate [12,13,28,29,47] have been observed. Many studies have also reported on deteriorating performance with increasing ambient relative humidity [12,13,28,34,49,51].

### Zeta-potential (of constituent materials)

The model postulates effective zeta-potential as one of the principal properties controlling the performance. Numerous studies have reported on determination of zeta-potential though often without providing any details on the method and conditions of the measurements (in particular, ionic composition and pH of solution) [12,24,28,29,34,47,49]. From the context, one can conclude that zeta-potentials were determined (probably using commercial instruments) from electrophoretic mobility (often in distilled water) of nanoparticles used to prepare the nanoporous films. Only in one study[48], it is explicitly stated that zeta-potential was determined from measurements performed with ELSZ-2000, Otsuka Electronics Co. instrument, and this presumably was done for assembled porous films (made of $MoS_2$ and/or $MoS_2+SiO_2$ in this case) in DI water. However, as discussed below, interpretation of electrokinetic measurements with porous films requires special procedures, which were not implemented in [48].

In most cases, zeta-potentials had noticeable magnitude but going beyond this qualitative observation is difficult because in nanoporous layers (effective) zeta-potentials could be quite different due to overlap of diffuse parts of EDLs as well as changes in the surface properties during the deposition processes. Besides, obtaining true zeta-potentials from electrophoretic mobility of nanoparticles (especially, in dilute electrolyte solutions) is non-trivial [53]. In view of this, in situ electrokinetic characterization of nanoporous films is advisable (see below).

### Nanoporous-film hydrophilicity

According to the model (see Eqs(33, 42,43)), the system performance is essentially controlled by the maximum capillary pressure, which increases with increasing hydrophilicity (see Eq(13)). In agreement with this, it was found experimentally that UV + $O_3$ treatment gave rise to smaller contact angles and strongly increased OCV [47,50], plasma treatment of carbon black dramatically decreased contact angle and increased OCV (ca.25 times) [12], plasma treatment of carbon nanofibers considerably improved performance [28] and air-plasma treatment of various solid oxides increased fully-wet length [29].

### Pore size

Our analysis reveals this as an important parameter because it controls the film hydraulic permeability (and its dependence on the pore size is quadratic) as well as maximum capillary pressure. However, amazingly little relevant information is provided in the literature. Only in [54] very limited information on the pore-size distribution (just one figure in the ESI) in nanoporous ZnO films was provided (obtained via interpretation of $N_2$ adsorption-desorption isotherms). In other studies, one has to rely on very rough "visual" inspection of electron micrographs. Thus for instance, the pores look like 200-300 nm large in [13,25] or 1 μm large in [49]. Sometimes, the size of "constituent" nanoparticles was determined but this is often not really conclusive especially with strongly anisometric particles. Thus for instance, the size of individual constituent carbon nanoparticles was 20-40 nm but in the assembled material "pores" look as large as ca. 100-200 nm [50]. Similarly, in [24] individual $V_2O_5$ nano-sheets have an apparent size of 100-300 nm but the roughly slit-like "pores" in the "restacked $V_2O_5$ membrane" appear to be rather about 1 μm large. This is not especially surprising given the pronounced anisometry of the constituent nano-sheets. In summary, direct in situ characterization of pore size in the nanoporous films is highly desirable.

### Film width

According to the model, the SCC should increase linearly with the film width while the OCV has to be independent of it. Exactly this was observed in several studies [24,29,34], sometimes the SCC increased somewhat super-linearly while the OCV increased slightly [28].

### Electrolyte concentration

The model predicts that at concentration-independent surface-charge density, both OCV and SCC should increase with decreasing electrolyte concentration and gradually tend to saturation. In some cases, experimental data qualitatively confirm this trend. Thus for instance, in [34] the OCV was approximately constant up to KCl concentration of 0.01 mM, then decreased about 3 times at 1 mM and dropped practically to zero in 100 mM KCl; in [12] up to 10 mM NaCl, the OCV decreased only about 2 times, then dropped dramatically; in [48] the OCV stayed practically constant (or even slightly increased with concentration up to ca. 1 mM, then dropped considerably; for films with approx. 1-µm pore size, the OCV decreased about 4 times when passing from DI water to tap water [49]. At the same time, as reported in [24], the SCC increased with NaCl concentration (up to 1 M) at shorter times (up to ca. 1000 s), while at longer times it somewhat decreased with it but not too much (< 2 times) and the dependence on concentration was non-monotone. This was observed for strongly anisotropic films probably having rather small pores oriented mostly along the film. In agreement with this picture of tangentially-oriented pores, evaporation in this case was unexpectedly slow. The time dependence might be due to a gradual salt accumulation in the film in the course of evaporation (the measurement started with pure water, then the device was immersed in electrolyte solutions of various concentrations).

The assumption of concentration-independent surface-charge density is probably not realistic because it can considerably decrease in dilute solutions (see, for example, [31,33]). This can give rise to non-monotone dependencies of power density on electrolyte concentration (which was observed, for example, in [48]) especially with smaller nanopores that are more promising due to larger maximum capillary pressures. Quantitative information on the surface-charge density can be obtained from in situ electrokinetic measurements (see below) combined with information on the pore size in the film.

### Film length and OCV distribution along the film

According to the model (Eq(35)), the SCC should increase linearly with the film length. This is due to the increasing evaporation area (and, thus, evaporated amount per unit time assuming area-independent linear evaporation rate, $q_e$). The OCV has to increase proportionally to the square of length (Eq(40)) because for longer films larger currents occur over longer distances. Qualitatively, these trends have been observed in several studies. Thus, the OCV increased slightly super-linearly with the film length although the SCC remained practically constant [24]. Both OCV and SCC increased with the film length (the voltage did so faster) although quantitatively the dependences don´t agree with the model [29]. Notably, the electrode stripes were broad (comparable to the spacing between their edges) in this case, which could influence the results.

In [34], both OCV and SCC increased strongly (the voltage still stronger than the current) when the film length increased from 1 cm to 2 cm, but for longer films the OCV practically didn´t change while the SCC went to zero. At the same time, the OCV was low at $L = 1\ cm$, which could be due to an external adhering water film giving rise to a "shortcut" of pressure gradients within this zone as argued in [25] (see also the next sub-section). The larger lengths might be above the fully-wet length but some conductance (needed for the OCV measurement) could still occur due to a broad pore size distribution and film anisotropy (see below about the limits of model applicability).

In [28], the SCC initially strongly increased with length, which could be due to the need to overcome an over-voltage of electrode reaction(s). This was followed by a decrease interpreted in terms of optimal layer length (ca. 6 cm in this case) and (implicit) assumption that the decrease occurred for "over-optimal" film lengths. Some SCC could still occur in this case due to the "diffuseness" of the boundary of the fully-wet zone caused by a pore-size distribution. The OCV also increased with the film length (and showed a maximum) but slower than the SCC. This is in qualitative agreement with the electron conductance of carbonized constituent nanofibers (see the next subsection).

*Distribution of OCV gradient along the film*

In some studies, sets of intermediate electrodes were fabricated that enabled observation of OCV-gradient distribution along the film. Our simple model predicts OCV gradients decreasing in a monotone way along the film (Eq(38)). This should occur because liquid is progressively lost to evaporation along the film, so ever smaller pressure gradients are needed to drive the remaining liquid along the film. Actually, an opposite (increasing) trend was often observed, at least, close to the film "entrance" followed by decreases starting at various positions along the film [12,25]. In [25], the increasing trend up to these positions was explained by adhering external water films (which were visually observed and proven to be longer with more hydrophilic film materials). These films reportedly gave rise to a "short-cut" of hydrostatic-pressure gradients inside the film, which also caused reduced OCV gradients (see Eq(7)). This was additionally confirmed by the fact that in films with increased hydrophilicity, the effect was more pronounced. Alternatively, a simple preliminary analysis in the ESI shows that the increasing trend followed by a decrease could also be a consequence of electron conductance of solid matrices of nanoporous-film materials (assuming ideal polarizability of their interfaces with the solution). Notably, in both studies [12,25], the matrices were carbon-based (so potentially electron-conducting). The OCV-gradient distribution has not been studied, yet, for materials with non-conduction matrices. This should be done in future studies.

## Film thickness

Typical nanoporous-film thicknesses ranged from tens to several hundred of micrometers, for example, 10 μm [51]; 16 μm [12,13]; 35 μm [24], 70-100 μm [50] or 100-300 μm [28]. In several studies films of various thicknesses were prepared and their performance compared. According to the model, for a constant film length, the SCC should be independent from the thickness because it is controlled by the rate of evaporation from the side surface (Eq(27)), and the OCV has to be inversely proportional to the thickness (Eq(33)) for with thicker films the same current flows through a medium with a lower electrical resistance. Besides, the fully-wet length should increase proportionally to the thickness (Eq(40)). In a qualitative agreement with the model, [25] observed that the fully-wet length was somewhat larger in thicker films while the OCV was somewhat lower. Similarly, the SCC was roughly independent of film thickness (though with a point "shooting up" at one thickness), with the OCV somewhat decreasing [28]. The OCV decreased with thickness (as it should) while the SCC increased, then decreased [49]. [24] reported that the SCC increased with the thickness up to its certain value, then decreased while the OCV increased. However, the material used in [24] (filtration-deposited $V_2O_5$ nano-flakes) was strongly anisotropic, which might explain the discrepancy. In [29], the SCC increased with thickness while the OCV remained approximately constant as if an unexpected increase of current were compensated by increasing conductance giving rise to a thickness-independent OCV.

In summary, correlations with the film thickness have often not been in agreement with the simple model, which probably can be explained by dependencies of specific (per thickness) properties of films on their thickness as well as by deviations from the assumptions of homogeneous and isotropic structure. Besides, correlations involving SCC might be affected by non-reversible electrode reactions.

## Film-matrix material

The standard electrokinetic model used above assumes the solid matrix to be non-conducting. Several studies on side-evaporation energy harvesting used such materials, for example, Ni-Al layered double hydroxides [51], AlOOH (nano-flakes)/UIO-66 (metal-organic nano-crystals) hybrid nanomaterials [34] or $Al_2O_3$ as well as a number of other oxides ($Fe_2O_3$, $Mn_3O_4$, ZnO, CuO, $SnO_2$, $Fe_3O_4$, $SiO_2$, $TiO_2$) [29]. Some of those oxides might have more or less pronounced semi-conducting properties but their impact was not explicit, and the performance was similar for all the oxides (including those certainly non-conducting like $Al_2O_3$ or $SiO_2$). The $V_2O_5$ nano-sheets used in [24] were also potentially semi-conducting but again the role of this factor was not clear. The composite nanostructures studied in [49] contained (possibly electron-conducting) carbon nano-spheres (CNSs) attached to $TiO_2$ nanowires but the CNSs probably did not form a continuous network. Notably, illumination of $TiO_2$-based porous films with UV light (giving rise to excitation of electron valence bands to conduction bands) gave rise to deteriorating performance in energy harvesting from evaporation [49].

At the same time, several studies used (potentially electron-conducting) carbon-based materials, namely, free-standing reduced graphene-oxide sponges [47], porous nanostructured carbon-black films [50] or plasma-treated carbon black [12]. [28] also used polymer-based nanofiber "mats" that became electron conducting after carbonization.

Thus, the role of electron conductance in this context remains unclear. Some studies speculated on the occurrence of "pseudo-streaming" currents [55], "electron concentration gradient driven by internal water flow in nanochannels…" [56], or that "ionic (or water molecular) motions …induce the charge carrier flows, the ionovoltaic phenomena, through Coulomb interactions (electron drag) at the solid/liquid interface" [54]. Ref.[48] clams that "the overcharged unipolar ions can drag their electric field during the electro-diffusion process, thereby inducing the forced flow of electrons in the semiconducting layer." Of course, ions "drag" their electric field but what exactly is meant by "overcharged" and why this should occur remains unclear.

So far, no quantitative analyses of these phenomena have been attempted for the systems of interest, and the mechanisms remain unclear. Coulomb-drag phenomena have been studied for very different systems (in particular, semiconductor multilayers or insulator-separated graphene sheets) [57], so the magnitude of (remotely) similar phenomena in the systems of interest cannot be estimated even by the order of magnitude. Meanwhile processes in systems involving two phases with different kinds of conductance (ionic in solution and electronic in the matrix) and interfaces between them are potentially quite complex, so their qualitative analysis (disregarding, for example, the essential electroneutrality requirement) can hardly be conclusive. In systems where this phenomenon has been investigated, the "secondary" currents have typically been much weaker than the "primary" ones. Drawing an analogy to the systems of interest, this would imply SCCs to be much smaller than the classical streaming currents. The estimates above were made assuming that ionic streaming currents can occur "in full". Much smaller "pseudo-streaming" currents would imply essentially lower power densities. Besides, one of the conclusions of the preliminary analysis of streaming potentials with electron-conducting substrates (carried out in the ESI) is that the conductivity can make the increase of streaming potential with pressure sublinear and essentially slower than in the case of non-conducting substrates. Along with the probably diminished SCCs induced via the Coulomb-drag mechanism, this can lead to still lower energy-harvesting efficiencies as compared with non-conducting substrates. Of course, in the latter case, some electrode reactions should occur, for example, water splitting.

## Future directions

### Model extensions

As we have just seen, predictions of our simple model do not always agree with experimental data. Given that the model assumptions have clear and simple physical meaning (discussed above), the probable reason for the discrepancies is deviations of properties of experimental systems from the <u>assumptions of a macroscopically homogeneous and isotropic porous film</u>. Besides, in systems with a broad pore-size distribution, the picture can be more complex, for example, because the partially-wet zone may become rather extended and diffuse. If the upper electrode is located within such a transition zone, dependencies on the film length may get essentially different from the predicted by the simple model. Besides, the average pore sizes controlling the maximum capillary pressure and the hydraulic permeability will be different. However, the aspects of film cross-sectional (in)homogeneity and pore-size distribution have not, yet, been addressed in the published studies, so hopefully the present analysis will stimulate corresponding experimental activities: for a more sophisticated modelling, additional experimental input is needed.

Assuming effectively <u>electrode reversibility</u> is a crude approximation. However, this is the only possible one at present due to the complete lack of published information on the electrochemical aspects of the process of interest. On the other hand, some over-voltages and kinetic limitation of electrode reactions definitely occur, which probably can explain some of the deviations of the observed trends from the model. These phenomena should be accounted for in future modelling studies. Postulating water-splitting as the electrode reaction, one should take into account the pH changes occurring on the electrodes and their possible impact on the surface-charge density.

As discussed above, the <u>role of porous-matrix electron/hole conductance</u> remains unclear. On one hand, similar phenomena have been observed for both non-conducting and (semi)conducting solid matrices of nanoporous films. On the other hand, the mechanisms of electrode reactions remain unclear especially in situations where the OCV was below 1.23 V (this was so in many cases) and water splitting could not occur. Occurrence of currents through (semi)conducting substrates driven by convective (counter)ion movement via the so-called Coulomb drag (or other mechanisms) appears imaginable but even order-of-magnitude estimates of their possible magnitude have not been performed, yet. This mechanism should be explored in future theoretical studies. They should also address other implications of substrate electron/hole conductivity for electrokinetic phenomena as outlined in the ESI.

### Experimental work

As demonstrated above, information on the electrokinetic properties of nanoporous films is of primary importance for the process of interest. Meanwhile, the corresponding experimental information is not really relevant because (effective) zeta-potentials were not determined in situ. In addition to effective zeta-potentials, pore-liquid electrical conductivity is also important (see Eqs(4,9,42)). Information on both these properties can be obtained directly for nanoporous films via measurements of tangential streaming current and potential in the so-called Adjustable-Gap Cell (AGC)[23]. In this arrangement, a pressure-driven tangential flow occurs along narrow (50-100 μm) slit-like gaps delimited by the investigated materials. Streaming potential and current are measured by a pair of reversible Ag/AgCl electrodes located close to the micro-channel edges. The system geometry and the electrode shape and location are such that genuine SCC (streaming current) can be determined. Moreover, the gap height can be finely tuned. From a series of measurements at several gap heights, one can estimate effective zeta-potential in the nanopores as well as solution electric conductivity in them [23]. Zeta-potential of the external film surface can also be obtained. This can afford some conclusions on the homogeneity of film properties. The measurements can be performed for a range of electrolyte concentrations and for various pH values.

Electrical conductance along nanoporous films was measured in several studies [12,24,54] but little information on the experimental conditions (DC/AC, frequency, applied voltage) has been provided. More systematic and rigorous studies are needed. They could complement the aforementioned studies of tangential electrokinetic phenomena.

From Eq(9), we see that besides the effective zeta-potential and electric conductivity, for estimates of EK-conversion efficiency, one needs to know the hydraulic permeability of nanoporous material. This cannot be obtained from the just outlined measurements in the commercially available AGC because for realistic minimum gap heights (ca.30 $\mu m$), the contribution of volume flow along the nanoporous layers is negligible compared to the flow through the gap itself. One can imagine dedicated designs with much smaller gap heights. However, to make the contribution of nanoporous films to the tangential volume flow noticeable, the gap height has to be in the range of single micrometers, which is probably comparable with the roughness of a major part of real nanoporous films. Alternatively, the hydraulic permeability of nanoporous films can be deduced from the dynamics of expansion (due to capillary impregnation) of wet annuli surrounding sessile drops placed on top of supported nanoporous films [58–60]. This can be done, in particular, directly with films used in systems for energy harvesting from side evaporation. True, for the interpretation in this case, one needs to know the maximum capillary pressure. This can be deduced from the maximum wet length provided that evaporation rate is measured in parallel (see Eq(33)).

As discussed above, the mechanisms of electrode reactions (which certainly should occur, at least, in systems with non-conducting solid matrices of nanoporous films) are completely unexplored in this context. Classical electrochemical methods (such as cyclic voltammetry, chrono-amperometry and/or chrono-potentiometry [61]) should be used. In the interpretation of such measurements, care has to be taken of the presence of nanoporous materials, which can influence the mass transfer phenomena and measurement interpretation.

To better understand the possible role of solid-matrix electron conductance, one should explore pressure-driven electrokinetic phenomena in media with ideally-polarizable electron-conducting matrices. On the other hand, the distribution of OCV (gradient) along porous films for non-conducting solid-matrix materials should also be studied to clarify if deviations from the model predictions can be due to other factors than the electron conductance.

As discussed above, average pore size and pore-size distribution are important properties because they control the hydraulic permeability of nanoporous films and maximum capillary pressure but little information on these important characteristics is available in the literature. Conventional as well as novel pore-size characterization techniques (for example, nitrogen, argon, water adsorption-desorption [62,63]) should be systematically used in situ for assembled nanoporous films.

## Conclusions

The efficiency of electromechanical energy conversion via electrokinetic phenomena is known to be not excessively high. Nevertheless, capillarity-driven electrokinetic electricity generation can be potentially rather efficient due to high capillary pressures in hydrophilic nanopores. This clearly manifests itself in the simple case of evaporation from single nanopores where power densities (per pore area) can theoretically reach several kW/m$^2$. This is due to the very fast evaporation from single-nanopore outlets occurring because of semi-circular vapor-diffusion pattern. Pressure drops of the order of 10 MPa (the maximum capillary pressures in 10-nm nanopores) can occur over pore lengths of several micrometers, which gives rise to large streaming potentials and streaming currents at the same time. However, with any appreciable density of nanopores in their arrays, the evaporation is controlled by vapor diffusion across stagnant layers in air whose typical thickness is around 1 mm. Although rather large streaming potentials can still occur, much lower streaming currents take place over much larger nanopore lengths required for the streaming potential to occur in full. Besides, this would require the use of layers of "monolith" nanomaterials with thicknesses in the range of millimeters.

Alternatively, there are configurations where evaporation area is much larger than the cross-section area of electrokinetic conversion. Two such configurations have been explored in the literature: an "in-series connection" of evaporation and EK-conversion elements and "side-evaporation devices" where evaporation and capillarity-induced volume flow occur "orthogonally". The latter configuration has recently been extensively explored experimentally though most studies have remained rather empirical. A simple model formulated in the present study has enabled analysis of correlations of system performance with the principal device characteristics such as the length, width and thickness of nanoporous layer, average pore size, hydrophilicity, effective zeta-potential or electric conductivity of nanopore liquid. Semi-quantitative comparison of these correlations with published experimental data reveals that some of the predicted correlations have been observed while others not. This may be related to the simplifying assumptions of the model, which has to be further developed to take into account, for example, pore-size distribution as well as nanoporous-layer cross-sectional inhomogeneity or anisotropy. Besides, our analysis of the literature reveals that several important kinds of experimental data are lacking. Primarily, this concerns information on effective zeta-potential and electric conductivity obtained in situ directly for assembled nanoporous films. Besides, experimental characterization of pore-size distribution is important. Finally, the mechanisms of electrode reactions and the role of electron/hole conductivity of solid matrices of nanoporous films remain largely unexplored. These issues should be investigated in future studies to help make the design of devices for energy harvesting from evaporation more rational and optimize their performance.

## Acknowledgements

The author acknowledges funding from the European Union through Project H2020-FETOPEN-2018-696 2019-2020-01-964524 "Energy harvesting via wetting/drying cycles with nanoporous electrodes (EHAWEDRY)". He is grateful to Dr. E.K. Zholkovskiy and Dr. A. Levchenko for fruitful discussions.

Electronic Supporting Information to:

# Evaporation-driven electrokinetic energy conversion: critical review, parametric analysis and perspectives

Andriy Yaroshchuk[a,b]

[a]ICREA, pg. L.Companys 23, 08010, Barcelona, Spain

[b]Department of Chemical Engineering, Universitat Politècnica de Catalunya, av. Diagonal 647, 08028, Barcelona, Spain

## Basics of electrokinetic phenomena in nanopores

Using the approach outlined in (Apel et al. 2021) and neglecting for simplicity salt-concentration gradients, from the transport equations derived in (Apel et al. 2021), we obtain this

$$J_v \left( \frac{1}{\chi} + \frac{\rho_{ek}^2}{g} \right) = -\nabla p + \frac{\rho_{ek}}{g} I \quad \text{(S1)}$$

$$-\nabla \varphi = \frac{I - \rho_{ek} \cdot J_v}{g} \quad \text{(S2)}$$

where $J_v$ is the volume flux, $p$ is the hydrostatic pressure, $\varphi$ is the electrostatic potential, $\chi$ is the hydraulic permeability at zero voltage gradient,

$$\rho_{ek} \equiv F(Z_1 \nu_1) \cdot c \cdot (\tau_1 - \tau_2) \quad \text{(S3)}$$

is the electrokinetic charge density (the proportionality coefficient between electric-current density and volume flux under streaming-current conditions, i.e. $\nabla c = 0, \nabla \varphi = 0$)

$$g \equiv \frac{F^2}{RT}(Z_1 \nu_1) c \left[ Z_1 \left( P_1 - \frac{\omega}{\nu_2} \right) - Z_2 \left( P_2 - \frac{\omega}{\nu_1} \right) \right] \quad \text{(S4)}$$

is the electric conductivity at zero transmembrane volume flow, $I$ is the electric-current density defined this way

$$I \equiv Z_1 J_1 + Z_2 J_2 \quad \text{(S5)}$$

Within the scope of model of straight capillaries, the coefficients featuring in Eqs(S1-S4) can be expressed this way

$$P_i \equiv \langle D_i \Gamma_i \rangle + RT c_i \langle F[1] \rangle \left( \frac{\langle \Gamma_i F[\Gamma_i] \rangle}{\langle F[1] \rangle} - \tau_i^2 \right) \quad \text{(S6)}$$

$$\omega \equiv RT \nu_1 \nu_2 c \langle F[1] \rangle \cdot \left( \frac{\langle \Gamma_1 F[\Gamma_2] \rangle}{\langle F[1] \rangle} - \tau_1 \tau_2 \right) \quad \text{(S7)}$$

$$\tau_i \equiv \frac{\langle \Gamma_i F[1] \rangle}{\langle F[1] \rangle} \quad \text{(S8)}$$

where $D_i$ are the ion diffusion coefficients, $\Gamma_i$ are the ion partitioning coefficients, $c$ is the virtual electrolyte concentration, $c_i \equiv \nu_i c$, $\nu_i$ are ion stoichiometric coefficients (they satisfy electroneutrality condition, $Z_1\nu_1 + Z_2\nu_2 = 0$), the brackets $\langle\ \rangle$ mean averaging over the pore cross section, $\hat{F}[\ ]$ is a linear functional operator giving a solution to this equation

$$\eta \nabla^2 \vec{v} = -\vec{f} \tag{S9}$$

where $\vec{f}$ is an arbitrary function of coordinate inside the pore. The form of operator $\hat{F}[\ ]$ depends on the pore geometry. For example, in long straight cylindrical pores of equal size, all the flows are 1D, besides, the ion partitioning coefficients, $\Gamma_i$, depend only on the radial coordinate inside the pore. The operator can be shown to have this form (Yaroshchuk and Bondarenko 2018)

$$\hat{F}[\Gamma_i] = -\frac{r_p^2}{\eta}\left[\ln(\rho)\int_0^\rho d\rho' \rho' \Gamma_i(\rho') + \int_\rho^1 d\rho' \rho' \ln(\rho') \Gamma_i(\rho')\right] \tag{S10}$$

where $\eta$ is the solution viscosity, $r_p$ is the pore radius, $\rho$ is the dimensionless radial coordinated scaled on the pore radius. The hydraulic permeability at zero voltage gradient in this case is equal to

$$\chi \equiv \langle F[1] \rangle = \frac{r_p^2}{8\eta} \tag{S11}$$

The ion transmission coefficients, $\tau_i$, defined by Eq(S8) quantify the extent to which ions are convectively entrained by the volume flow. Notably, these coefficients are larger than one for counterions whose partitioning coefficients exceed unity. In principle, these coefficients can be affected by steric hindrance(Yaroshchuk et al. 2019) but this is not significant in nanopores whose size is much larger than the ion size (the focus of this study). Based on the same considerations, we also neglect the effect of steric hindrance on the ion diffusion and consider ion diffusion coefficients in nanopores constant and equal to those in bulk electrolyte solution.

Popular space-charge model postulates ion partitioning due to electrostatic interactions with fixed charges on the nanopore walls and local thermodynamic equilibrium (Yaroshchuk 2011). Accordingly, ion-partitioning coefficients can be obtained from the condition of constant electrochemical potential for each ion across the nanopore cross-section (Boltzmann distribution)

$$\Gamma_i = exp(-Z_i \psi) \tag{S12}$$

In combination with Poisson equation, this gives rise to Poisson-Boltzmann (PB) equation for the quasi-equilibrium dimensionless electrostatic potential, $\psi$

$$\nabla^2 \psi = \frac{(\kappa r_p)^2}{Z_1 - Z_2}\left(exp(-Z_1\psi) - exp(-Z_2\psi)\right) \tag{S13}$$

where $\kappa$ is the reciprocal Debye screening length defined as

$$\kappa \equiv \sqrt{\frac{2F^2 I_f}{\varepsilon\varepsilon_0 RT}} \tag{S14}$$

$$I_f \equiv \frac{1}{2} Z_1 \nu_1 (Z_1 - Z_2) \cdot c \tag{S15}$$

is the ionic strength in the virtual solution. For the cylindrical pore geometry, the boundary conditions are zero potential derivative at the pore axis (from the symmetry) and a given electric-charge density (potential derivative) at the capillary wall.

$$\frac{d\psi}{d\rho}\bigg|_{\rho=1} = \frac{F\sigma r_p}{\varepsilon\varepsilon_0 RT} \tag{S16}$$

where $\sigma$ is the surface-charge density. One can also consider the so-called charge-regulation boundary condition (Israelachvili 2011). PB equation has several approximate solutions but they have limited applicability. Therefore, PB equation will be solved numerically. The integrations featuring in Eqs(S6-S8) will be performed numerically, too. See the ESI of (Yaroshchuk and Bondarenko 2018) for more detail on the procedures.

# Nanopore conductivity (straight cylindrical nanopores, (1:1) electrolytes)

It has been shown that the second term in the right-hand side of Eq(S6) as well as the whole "mutual electro-diffusion" term given be Eq(S7) are very small in nanopores with well overlapped diffuse parts of electric double layers and are limited from above by 10-20% in broader nanopores (Yaroshchuk 1995). Therefore, for approximate estimates these terms can be neglected. Substituting the corresponding simplified version of Eq(S6) to Eq(S4), assuming constant diffusion coefficients, (1:1) electrolyte, performing cross-section averaging for cylindrical pore geometry, and using Eq(S12) for the ion-partitioning coefficients, we obtain

$$g = \frac{F^2 c}{RT} \frac{2}{a^2} \int_0^a \left( D_+ exp\left(-\frac{F\psi}{RT}\right) + D_- exp\left(\frac{F\psi}{RT}\right) \right) r dr \tag{S17}$$

Estimates of streaming potential and EK-conversion efficiency for nanoporous films with pore size and surface charge mimicking polymer track-etched membranes

The surface-charge density in nanopores (24 nm diameter) of poly-ethylene-terephtalate track-etched membranes has recently been determined from simultaneous measurements of osmotic pressure and salt diffusion in KCl and LiCl solutions (Apel et al. 2021). For the surface-charge densities fitted to experimental data in this study (and identical cylindrical pores of 24-nm diameter), one calculate numerically (using the procedures described in (Yaroshchuk and Bondarenko 2018)) the electrokinetic-charge density, electric conductivity and hydraulic permeability at zero current (inverse of coefficient by volume flux in Eq(S1)) as well as streaming potential at maximum capillary pressure at complete wetting (ca.1.2 MPa) and EK-conversion efficiency for the side-evaporation configuration. The results are listed in Table S1. The calculations assumed the real porosity of track-etched membranes (ca.3.6%) but from the definitions of streaming-potential coefficient (Eq(43)) and EK-conversion efficiency (Eq(9)) one can see that these properties are independent of porosity because electrokinetic-charge density is independent of porosity (see Eq(S3)) while both hydraulic permeability and electric conductivity in the definition of EK-conversion efficiency are directly proportional to it.

Table S1. Streaming potential and EK-conversion efficiency in nanoporous materials mimicking electro-surface properties of track-etched membranes studied in (Apel et al. 2021)

| Salt/ concentration | surface-charge density $(mA \cdot s/m^2)$ | electrokinetic-charge density $(A \cdot s/dm^3)$ | electric conductivity $(mS/m)$ | hydraulic permeability $(nm/(s \cdot MPa))$ | streaming potential (V) | electrokinetic-conversion efficiency $(16\theta/27)$ (-) |
|---|---|---|---|---|---|---|
| KCl, 1.5 mM | -5.7 | -73 | 2.8 | 0.64 | -1.9 | 0.018 |
| KCl, 3 mM | -9.5 | -107 | 4.9 | 0.62 | -1.6 | 0.022 |
| LiCl, 1.5 mM | -5.6 | -72 | 1.4 | 0.58 | -3.4 | 0.031 |
| LiCl, 3 mM | -9.3 | -106 | 2.5 | 0.55 | -2.7 | 0.036 |

Remarkably, while the streaming potential decreases with electrolyte concentration, the EK-conversion efficiency even somewhat increases. This is due to the considerable increase of surface-charge (and electrokinetic-charge) density with the electrolyte concentration.

## Electrokinetics with electron-conducting substrates

In this section, we make an attempt of taking into account electron/hole conductance of matrix of nanoporous materials experiencing electrokinetic phenomena (streaming potential). We consider the simplest limiting case of sufficiently large pores without any appreciable overlap of diffuse parts of electric double layers.

## Pressure-driven mode

In this section, we consider long straight channels with a pressure-drive flow. According to Eqs(4,7),

$$\frac{d\varphi}{dx} = \frac{\varepsilon\varepsilon_0}{\eta g}(\zeta - \bar{\psi})\frac{dP}{dx} \tag{S18}$$

In sufficiently broad channels, the average electrostatic potential, $\bar{\psi}$, is very small and can be neglected. Therefore, from Eq(S18), we obtain

$$d\varphi = \frac{\varepsilon\varepsilon_0 \zeta}{\eta g} dP \tag{S19}$$

Here, $\varphi$ is the electrostatic potential in the central part of the channel far away from its surfaces. Besides it, close to the surfaces there is a potential drop within diffuse part of electric double layer (zeta-potential). With electron-conducting substrates, the electrostatic potential of conductor surface must be the same all the way along the channel. In particular, this surface potential occurs at zero volume flow due to preferential adsorption of ions of one sign on the surface (or dissociation of ionogenic groups). Let us denote this constant surface potential $\zeta_0$, so

$$\zeta + \varphi = \zeta_0 \tag{S20}$$

By substituting Eq(S20) to Eq(S19), we obtain

$$\frac{d\varphi}{\zeta_0 - \varphi} = \alpha dP \tag{S21}$$

where we have denoted

$$\alpha \equiv \frac{\varepsilon \varepsilon_0}{\eta g} \tag{S22}$$

In this simple analysis, we neglect the influence of electrokinetic phenomena on the volume flow (this is justified just in sufficiently broad channels). Therefore, the hydrostatic pressure is independent of electrostatic potential and its profile is linear. Accordingly, Eq(S21) can be easily integrated along the channel to yield

$$\alpha \Delta P = \ln\left(\frac{\zeta_0 - \varphi(L)}{\zeta_0 - \varphi(0)}\right) \tag{S23}$$

where $L$ is the channel length,

$$\Delta P \equiv P(0) - P(L) \tag{S24}$$

is the hydrostatic-pressure difference along the channel. From Eqs(S23, S24) we obtain

$$\Delta \varphi \equiv \varphi(0) - \varphi(L) = (\zeta_0 - \varphi(0))(exp(\alpha \Delta P) - 1) \tag{S25}$$

The hydrostatic-pressure profile is linear

$$P(x) = P(0) - \frac{\Delta P}{L} x \tag{S26}$$

Taking this into account, from Eq(S21), we obtain

$$\varphi(x) - \varphi(0) = (\zeta_0 - \varphi(0))\left(1 - exp\left(\alpha \Delta P \frac{x}{L}\right)\right) \tag{S27}$$

One can see that, in contrast to the classical case of dielectric substrates, the electrostatic-potential profile is non-linear. The extent of non-linearity is controlled by the value of parameter $\alpha \Delta P$. Notably, when this parameter is small, we recover the same linear potential profile as in the classical case.

Physically, the constancy of surface electrostatic potential is ensured due to the appearance of polarization electron charges at the channel surfaces. Together with the initially present fixed charges (arising due to preferential ion adsorption or dissociation of ionogenic groups) they give rise to a position-dependent zeta-potential that can be found from the condition of constancy of surface electrostatic potential (Eq(S20)) and distribution of electrostatic potential outside the EDLs (Eq(S27))

$$\zeta(\xi) \equiv \zeta_0 - \varphi(\xi) \equiv (\zeta_0 - \varphi(0)) exp(\alpha \Delta P \xi) \tag{S28}$$

where $\xi \equiv x/L$ is the dimensionless coordinate scaled on the cannel length. We consider the conductor ungrounded. Therefore, the total induced electron charge is zero. There is this well-known relationship between surface-charge density and equilibrium electrostatic potential at a charged surface (Gouy-Chapman formula)

$$\sigma = 2\sqrt{2RT\varepsilon\varepsilon_0 c} \cdot sinh\left(\frac{\zeta}{2}\right) \tag{S29}$$

The surface-charge density is proportional to the hyperbolic sinus of zeta-potential. Taking into account this and the fact that the total surface charge under flow conditions must be equal to the charge at no flow, we obtain

$$sinh\left(\frac{\zeta_0}{2}\right) = \int_0^{L1} sinh\left(\frac{\zeta(\xi)}{2}\right) d\xi \tag{S30}$$

Substituting Eq(S28) for the distribution of zeta-potential, we obtain

$$sinh\left(\frac{\zeta_0}{2}\right) = \int_0^1 sinh\left(\left(\frac{\zeta_0-\varphi(0)}{2}\right)exp(\alpha\Delta P\xi)\right)d\xi \tag{S31}$$

The integral in the right-hand side of Eq(S31) can be taken in terms of special functions

$$A \cdot sinh\left(\frac{\zeta_0}{2}\right) = Shi\left(\left(\frac{\zeta_0-\varphi(0)}{2}\right)e^A\right) - Shi\left(\frac{\zeta_0-\varphi(0)}{2}\right) \tag{S32}$$

where $Shi$ is the integral hyperbolic sinus, and we have denoted

$$A \equiv \alpha\Delta P \tag{S33}$$

Eq(S32) is a transcendental equation for the determination of "entrance" electrostatic potential, $\varphi(0)$, as a function of parameter $A$, which is proportional to the hydrostatic-pressure drop along the channel. Given that integral hyperbolic sinus is a strongly increasing function of its argument, Eq(S32) shows that when parameter $A$ increases, $\zeta_0 - \varphi(0) \to 0$. Physically, this means that the polarization charge distributes itself in such a way that the net charge density (fixed plus induced charge) at the channel "entrance" tends to zero whereas it "peaks" exponentially ever stronger (with increasing pressure difference) close to the "exit" (see Eq(S28)). Fig.S1 shows that for conducting substrates the dependence of streaming potential on applied pressure is essentially sublinear. This can be a problem in view of achieving sufficiently high streaming potentials in energy harvesting from evaporation.

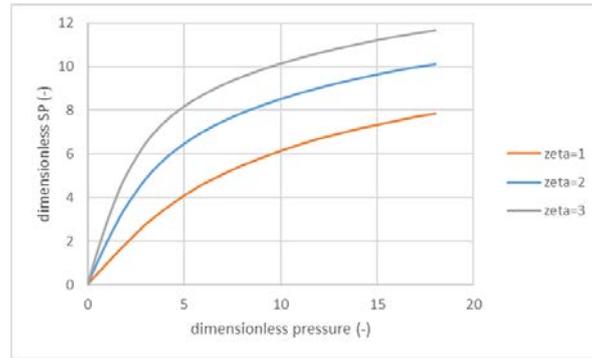

Fig.S1. Dimensionless OCV (streaming potential) as a function of dimensionless hydrostatic-pressure difference (parameter $A$ defined by Eq(S33)); the values of "equilibrium" dimensionless zeta-potential ($\zeta_0$) are indicated in the legend.

### Evaporation-driven mode

We consider the system with side evaporation schematically shown in Fig.7. In this case, Eq(S18) is still applicable but the hydrostatic-pressure gradient is not constant but is given by Eq(30). Therefore, by assuming $\bar{\psi} = 0$ and using Eq(S20), we obtain

$$\frac{d\varphi}{\zeta_0-\varphi} = \frac{\alpha q_e}{\chi h}(x-L)dx \tag{S34}$$

This can be easily integrated to yield

$$-ln\left(\frac{\zeta_0-\varphi(x)}{\zeta_0-\varphi(0)}\right) = \frac{\alpha q_e}{\chi h}x\left(\frac{x}{2}-L\right) \tag{S35}$$

$$\varphi(\xi) - \varphi(0) = (\zeta_0 - \varphi(0))\left(1 - exp\left(\beta\xi\left(1-\frac{\xi}{2}\right)1\right)\right) \tag{S36}$$

where $\xi \equiv x/L$ is the dimensionless coordinate along the porous film,

$$\beta \equiv \frac{\varepsilon\varepsilon_0}{\eta g}\frac{q_e L^2}{\chi h} \tag{S37}$$

The local zeta-potential, $\zeta(\xi) \equiv \zeta_0 - \varphi(\xi)$, is

$$\zeta(\xi) = (\zeta_0 - \varphi(0))exp\left(\beta\xi\left(1-\frac{\xi}{2}\right)\right) \tag{S38}$$

Taking into account as previously that the total induced electric charge should be zero and using Eq(S29), we obtain.

$$sinh\left(\frac{\zeta_0}{2}\right) = \int_0^1 sinh\left(\left(\frac{\zeta_0-\varphi(0)}{2}\right)exp\left(\beta\xi\left(1-\frac{\xi}{2}\right)\right)\right)d\xi \tag{S39}$$

From this transcendental equation, one can find $\varphi(0)$ as a function of $\beta$. From Eq(S36) we obtain this expression for the coordinate dependence of derivative of electrostatic potential

$$\frac{d\varphi}{d\xi} = -(\zeta_0 - \varphi(0))\beta(1-\xi)exp\left(\beta\xi\left(1-\frac{\xi}{2}\right)\right) \tag{S40}$$

Fig.S2 shows an example of distribution of electrostatic-potential derivative.

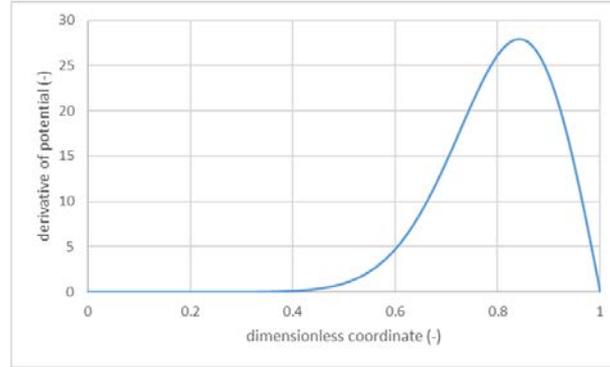

Fig.S2. Distribution of derivative of dimensionless OCV (streaming potential) with respect to dimensionless coordinate along porous film calculated using Eq(S40): $\zeta_0 = -3, \beta = 40$

This distribution is in qualitative agreement with experimental data obtained in (Xue et al. 2017) for nanoporous films made from carbon-black nanoparticles.